\documentclass[prx,showpacs,superscriptaddress,floatfix,twocolumn]{revtex4-2}

\usepackage[]{graphicx}
\usepackage{tabularx}
\usepackage[usenames,dvipsnames]{color}
\usepackage{soul}
\usepackage{bm}
\usepackage{amsmath}
\usepackage{lineno}
\usepackage{gensymb}

\usepackage{times}
\usepackage{amsfonts}
\usepackage{mathrsfs}
\usepackage{graphicx}
\usepackage{dcolumn}
\usepackage{bm}
\usepackage{color}

\usepackage[colorlinks,bookmarks=false,citecolor=blue,linkcolor=red,urlcolor=blue]{hyperref}
\usepackage{multirow}
\usepackage{physics}

\newcommand\FM[1]{\textcolor{black}{ #1}}
\newcommand\MC[1]{\textcolor{black}{ #1}}

\begin{document}

\title{{Signatures of a} surface spin-orbital chiral metal}

\author{Federico Mazzola**}
\email{federico.mazzola@unive.it}
\affiliation{Department of Molecular Sciences and Nanosystems, Ca’ Foscari University of Venice, Venice IT-30172, Italy}
\affiliation{Istituto Officina dei Materiali, Consiglio Nazionale delle Ricerche, Trieste I-34149, Italy}

\author{Wojciech Brzezicki**}
\email{w.brzezicki@uj.edu.pl}
\affiliation{Institute of Theoretical Physics, Jagiellonian University, ulic, S. \L{}ojasiewicza 11, PL-30348 Krak\'ow, Poland}
\affiliation{International Research Centre MagTop, Institute of Physics, Polish Academy of Sciences, Aleja Lotnik\'ow 32/46, PL-02668 Warsaw, Poland}

\author{Maria Teresa Mercaldo}
\affiliation{Dipartimento di Fisica ``E. R. Caianiello", Universit\`a di Salerno, IT-84084 Fisciano (SA), Italy}

\author{Anita Guarino}
\affiliation{CNR-SPIN, c/o Universit\'a di Salerno, IT-84084 Fisciano (SA), Italy}

\author{Chiara Bigi}
\affiliation{Synchrotron SOLEIL, F-91190 Saint-Aubin, France}

\author{Jill A. Miwa}
\affiliation{Department of Physics and Astronomy, Interdisciplinary Nanoscience Center, Aarhus University, 8000 Aarhus C, Denmark}

\author{Domenico De Fazio}
\affiliation{Department of Molecular Sciences and Nanosystems, Ca’ Foscari University of Venice, Venice IT-30172, Italy}

\author{Alberto Crepaldi}
\affiliation{Dipartimento di Fisica, Politecnico di Milano, Piazza Leonardo Da Vinci 32, Milano IT-20133, Italy}

\author{Jun Fujii}
\affiliation{Istituto Officina dei Materiali, Consiglio Nazionale delle Ricerche, Trieste IT-34149, Italy}

\author{Giorgio Rossi}
\affiliation{Dipartimento di Fisica, Università degli studi di Milano, IT-20133 Milano, Italy}
\affiliation{Istituto Officina dei Materiali, Consiglio Nazionale delle Ricerche, Trieste IT-34149, Italy}

\author{Pasquale Orgiani}
\affiliation{Istituto Officina dei Materiali, Consiglio Nazionale delle Ricerche, Trieste IT-34149, Italy}

\author{Sandeep Kumar Chaluvadi}
\affiliation{Istituto Officina dei Materiali, Consiglio Nazionale delle Ricerche, Trieste IT-34149, Italy}

\author{ Shyni Punathum Chalil}
\affiliation{Istituto Officina dei Materiali, Consiglio Nazionale delle Ricerche, Trieste IT-34149, Italy}

\author{Giancarlo Panaccione}
\affiliation{Istituto Officina dei Materiali, Consiglio Nazionale delle Ricerche, Trieste IT-34149, Italy}

\author{Anupam Jana}
\affiliation{Istituto Officina dei Materiali, Consiglio Nazionale delle Ricerche, Trieste IT-34149, Italy}

\author{Vincent Polewczyk}
\affiliation{Istituto Officina dei Materiali, Consiglio Nazionale delle Ricerche, Trieste IT-34149, Italy}

\author{Ivana Vobornik}
\affiliation{Istituto Officina dei Materiali, Consiglio Nazionale delle Ricerche, Trieste IT-34149, Italy}

\author{Changyoung Kim}
\affiliation{Department of Physics and Astronomy, Seoul National University, Seoul, 08826, Korea}

\author{Fabio~Miletto Granozio}
\affiliation{CNR-SPIN, c/o Complesso di Monte S. Angelo, IT-80126 Napoli, Italy}

\author{Rosalba Fittipaldi}
\affiliation{CNR-SPIN, c/o Universit\'a di Salerno, IT-84084 Fisciano (SA), Italy}

\author{Carmine Ortix}
\affiliation{Dipartimento di Fisica ``E. R. Caianiello", Universit\`a di Salerno, IT-84084 Fisciano (SA), Italy}

\author{Mario Cuoco}\email{mario.cuoco@spin.cnr.it}
\affiliation{CNR-SPIN, c/o Universit\'a di Salerno, IT-84084 Fisciano (SA), Italy}

\author{Antonio Vecchione}\email{antonio.vecchione@spin.cnr.it}
\affiliation{CNR-SPIN, c/o Universit\'a di Salerno, IT-84084 Fisciano (SA), Italy}

\begin{abstract}
{The 
\FM{relation}
between crystal symmetries, electron correlations, and electronic structure 
steers 
the formation of a large array of unconventional phases of matter, 
\FM{including}
magneto-electric loop currents and chiral magnetism \cite{Xu_2022, Grytsiuk_2020, Zhang_2022, Barron_2000, Bode_2007, Cheong_2022}. \MC{Detection of such hidden 
orders
is a major goal in condensed matter physics}. However, to date, 
nonstandard forms of magnetism with chiral electronic ordering have been experimentally elusive \cite{Guo_2022}.
\FM{Here, we develop a theory for symmetry-broken chiral ground states and propose a methodology based on circularly polarized spin-selective angular-resolved photoelectron spectroscopy to probe them.}
We exploit the archetypal \MC{quantum material} Sr$_2$RuO$_4$ and \FM{reveal spectroscopic signatures which, even though subtle, {may} be reconciled} with the formation of spin-orbital \MC{chiral}
currents at the material surface \cite{Fittipaldi_2021, Imai_1998, Grinenko_2020}. As we shed light on these 
chiral regimes, our findings pave the way for a deeper understanding of ordering phenomena and unconventional magnetism.}
\end{abstract}

\maketitle
\noindent
\noindent{\bf{\large Main}}\\
One central problem in condensed matter physics is the existence of unconventional magnetism
beyond the traditional forms, arising from the long-range order of magnetic dipole moments $\mu=(2 S+L)${,} with $S$ and $L$ being the electron spin and orbital angular momentum, respectively \cite{Mishra_2020, Kao_2022, Jiun-Haw_2012, Wang_2015, Zhao_2016}. Most often, such magnetic dipole moments do not arrange spatially with ordered patterns in the crystal but other forms of magnetic phases may still originate from an electronic ordering made of charge currents at the atomic scale.
Such phases are odd in time (time-reversal symmetry is broken), \FM{inherently subtle and difficult to observe}, and often associated with a hidden magnetic order \cite{Xu_2022, Bounoua_2020, Grytsiuk_2020}.
 
\begin{figure*}[!t]
\centering
\includegraphics[width=0.7\textwidth,angle=0,clip=true]{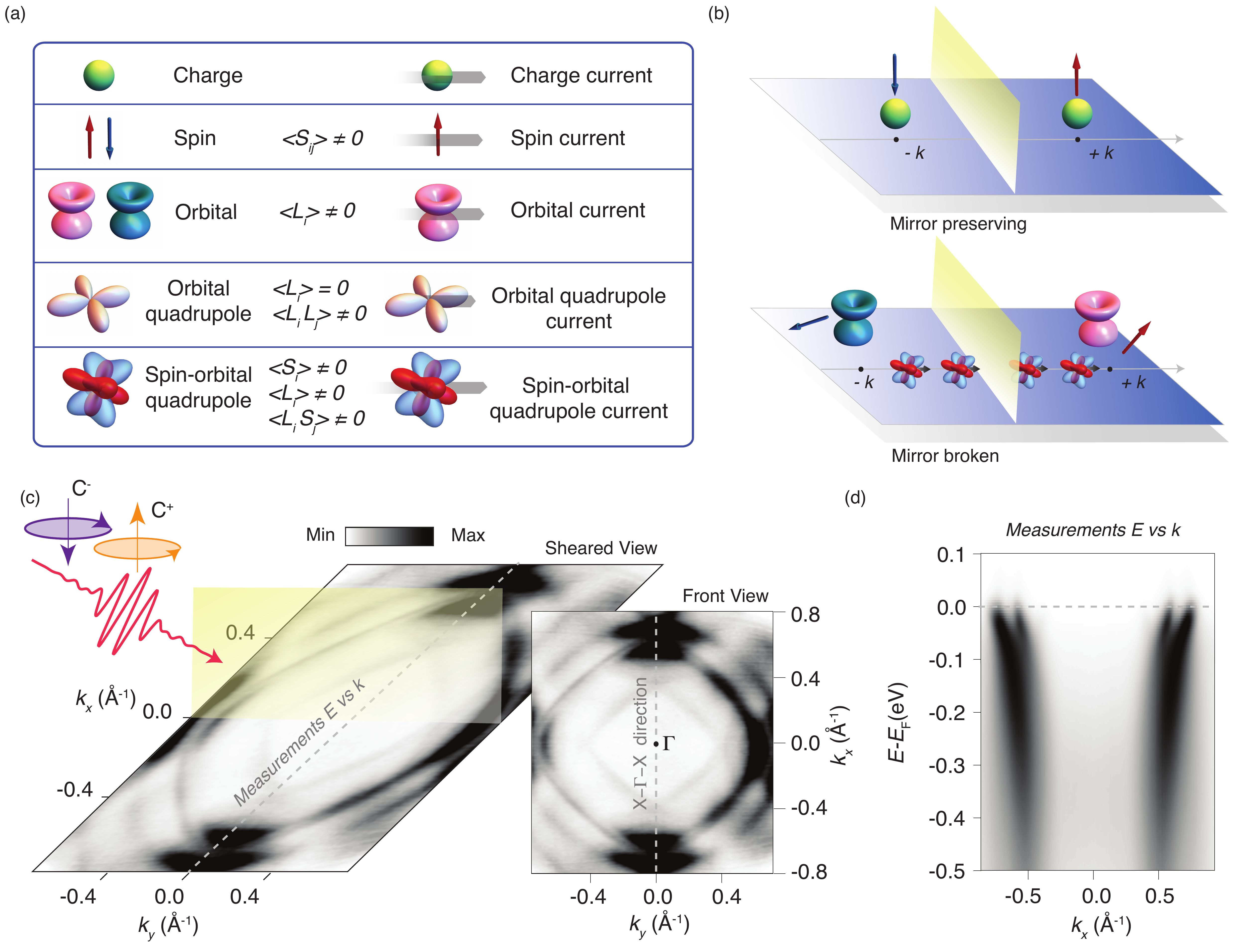}\caption{{\bf Currents and symmetries in an \MC{electronic} system.} {\bf a} Legend of possible charge, spin and orbital currents that can be created in a material. As the charge can give rise through its motion to a conventional current, i.e. charge current, spin and orbital dipoles or quadrupoles can also generate more complex types of currents. Indeed, the spin and orbital angular momentum are pseudo-vectors that change sign upon time-reversal symmetry, and then the spin and orbital currents carrying spin or orbital dipoles are time-reversal conserving. Instead, the currents carrying orbital or spin-orbital quadrupoles break time-reversal symmetry and yield non-vanishing amplitudes for the dipole and quadrupole observables at given momentum. 
{\bf b} Examples of (top) mirror preserving and (bottom) mirror broken configurations. In a system which preserves time-reversal, a charge with its spin at a certain positive momentum, under the action of such symmetries, goes into a charge with opposite spin (directed still along the same direction but opposite in sign) at negative (symmetry related) momentum. For mirror-symmetric configurations the sign changes occurs when the spin lies in the mirror plane as depicted in the sketch. The situation is different when a spin-orbital quadrupolar current, which flows along a given direction is included. Such current couples to the orbital ($L$) and spin ($S$) angular momentum, creating a strong asymmetry in their product $LS$ (quadrupole) between symmetry-related momenta in the Brillouin zone. This is why the spin is not longer reversed and picks up an angle. {\bf c} Experimental configuration adopted to measure the asymmetry of $LS$ caused by the chiral current-driven breaking of mirror symmetry. The Fermi surface (shown in both sheared and front view for clarity) of Sr$_2$RuO$_4$, employed here as a test-bed for our theory. {\bf d} Energy of the electrons as a function of momentum collected by photoelectron spectroscopy for Sr$_2$RuO$_4$ along the direction indicated in {\bf c} by the dashed grey line.}\label{fig1}
\end{figure*}

The transport properties of all metallic quantum materials are determined by the spin and orbital degrees of freedom of the Fermi surface. Broken symmetry states (mirror and/or time) with charge currents can have an internal structure with a combination of spin and orbital angular momentum. Since the spin and orbital angular momentum are pseudovectors with magnetic dipolar nature, one can refer to their product as a spin-orbital quadrupole or orbital quadrupole. Spin and orbital angular momentum \MC{are odd in time} and the current is an odd function of crystal wave vector. Therefore, \MC{currents carrying spin-orbital or orbital quadrupoles, which are even in time, break time-reversal symmetry (Fig.\ref{fig1}{\bf{a}}) \cite{Fittipaldi_2021, Tang_2023, Zapf_2023, Liu_2019}}.

A hallmark of both orbital and spin-orbital quadrupole current is the appearance of additional symmetry breaking related to mirror, inversion, or roto-inversion transformations, more generically known as chirality. Hence, a chiral electronic ordering may be realized uniquely as a consequence of the intrinsic spin and orbital structure of the charge currents \cite{Hasan_2021, Xin_2020, Chen_2021}. Chirality is known to set out several unconventional forms of transport and magnetism \cite{Grytsiuk_2020, Avci_2019, Casher_1974, Fukushima_2008, Rikken_2022, Legg_2022, Atzori_2021}. \FM{However, chiral effects have been proven challenging to be detected, because {their electronic signature is} very weak}. So far, their measurements are limited to only a few material-specific cases \cite{Xu_2022, Barron_2008, Yang_2022, Shekhar_2018}.


In a metallic state, chiral orders imprint {themselves on} the spin and orbital textures of the electronic states close to the Fermi level \cite{Train_2008, Ma_2017, Li_2016}. The action of mirror- and time-reversal symmetries connects the amplitude of spin- and orbital-angular momentum of the electron states at symmetry-related momenta. For instance, for time-reversal symmetric electronic states (Fig.\ref{fig1}{\bf b}), the associated spin-angular momentum must have opposite orientations, i.e. $(-\bm{k}, \uparrow)$ transforms into $(+\bm{k}, \downarrow)$ \cite{Schuler_2020}. The same behaviour applies to the orbital-angular momentum  $L$. On the other hand, for mirror-symmetric electronic states, due to the axial nature of $S$ and $L$, upon mirror transformation the components lying within the mirror plane change sign, while the ones perpendicular to the plane remain unchanged. 
Apart from the dipolar one, the interaction between spin and orbital degrees of freedom can set out physical observable{s} with tensorial character. Here, the time-reversal and mirror symmetries are also expected to affect the behavior of the spin-orbital ($L_i S_k$) and orbital ($L_i L_k$) quadrupole components, when probed at symmetry-related momenta. This means that the spin-orbital dipolar and quadrupolar structures are the relevant observables of the onset of a symmetry breaking and can be exploited to assess the nature of the realized electronic ordering. For a ground state hosting chiral currents, there is an antisymmetric combination of $L$ and $S$. In this situation, the spin-orbital texture of the electronic state{s} at the Fermi level {exhibit} a distinctive behavior: \FM{Chiral currents carrying spin-orbital quadrupole{s} give rise to orbital moments with the same parity as for mirror-symmetric systems{,} while spin-orbital quadrupoles {themselves} show neither {a} time nor mirror-symmetric profile. {It is this physical case, that we focus on in this work.}}

To exemplify the concept of {a} chiral metal, one can employ the analogy with chiral crystals and their symmetry properties. \MC{One can then generally identify a chiral metal with an electronic state that has a well-defined handedness, due to the lack of inversion, mirror or other roto-inversion symmetries \cite{Moss1996,Flack2003}.}
{In this study, we start from this description to introduce the concept of surface spin-orbital chiral metal to \FM{indicate} a conducting electronic state of matter that has a well-defined handedness due to an interaction driven by magnetochiral order that lacks mirror symmetries, resulting from the internal spin-orbital structure, {\FM but} possesses the same translational symmetry as the hosting crystal.}

Here, we \FM{show} the relationship between the spin-orbital textures of the electronic states, for both dipolar and quadrupolar channels, and the consequential occurrence of a chiral electronic ordering. By supporting the theory with circularly polarized spin-selective angular-resolved photoelectron spectroscopy, we \FM{introduce a methodology to probe, otherwise elusive, 
\MC{symmetry-broken chiral electronic states.}
} To this aim, we exploit the archetypal quantum material Sr$_2$RuO$_4$ (see Methods for growth and measurements details) and reveal \FM{signatures of} a broken symmetry phase \cite{Fittipaldi_2021}, compatible with the formation of spin-orbital quadrupole currents at the material surface.
\\

\noindent{\bf{\large Dichroic and spin-dichroic photoemission effects}}\\
As anticipated, assessing whether the profile of the $LS$ quadrupole components has a mirror-and time-broken character is key for the detection of electronic phases with electronic chiral currents. 
\MC{Sr$_2$RuO$_4$ is an ideal candidate to host symmetry-broken chiral ground states because of the recent low-energy muon spin spectroscopy \cite{Fittipaldi_2021} and scanning tunnelling microscopy (STM) measurements \cite{Marques_2021}, which unveil the existence of unconventional magnetism and electronic ordering forming at the surface.}

To date, nevertheless, there are no accepted methodologies to ascertain the existence of such \MC{symmetry-broken chiral states}. From our theory, in order to detect the action of symmetries on $LS$, one has to probe the out-of-plane components of the orbital angular-momentum ($L_{z}$) and {spin} ($S_z$). Experimentally, the $LS$ asymmetry can be tackled by circularly polarized spin-selective angular-resolved photoelectron spectroscopy (CP-Spin-ARPES) \cite{Schuler_2020, DiSante_2023}. \FM{This approach requires extreme caution in the alignment and geometry of the apparatus (Fig.\ref{fig1}c). Indeed, photoelectrons from circularly polarized light host a combination of intrinsic and geometric matrix elements \cite{Razzoli_2017, Cho_2018}. Nevertheless, as shown in Ref.\cite{Cho_2018}, these can be disentangled (see Methods).}


\begin{figure*}[!t]
\centering
\includegraphics[width=0.6\textwidth,angle=0,clip=true]{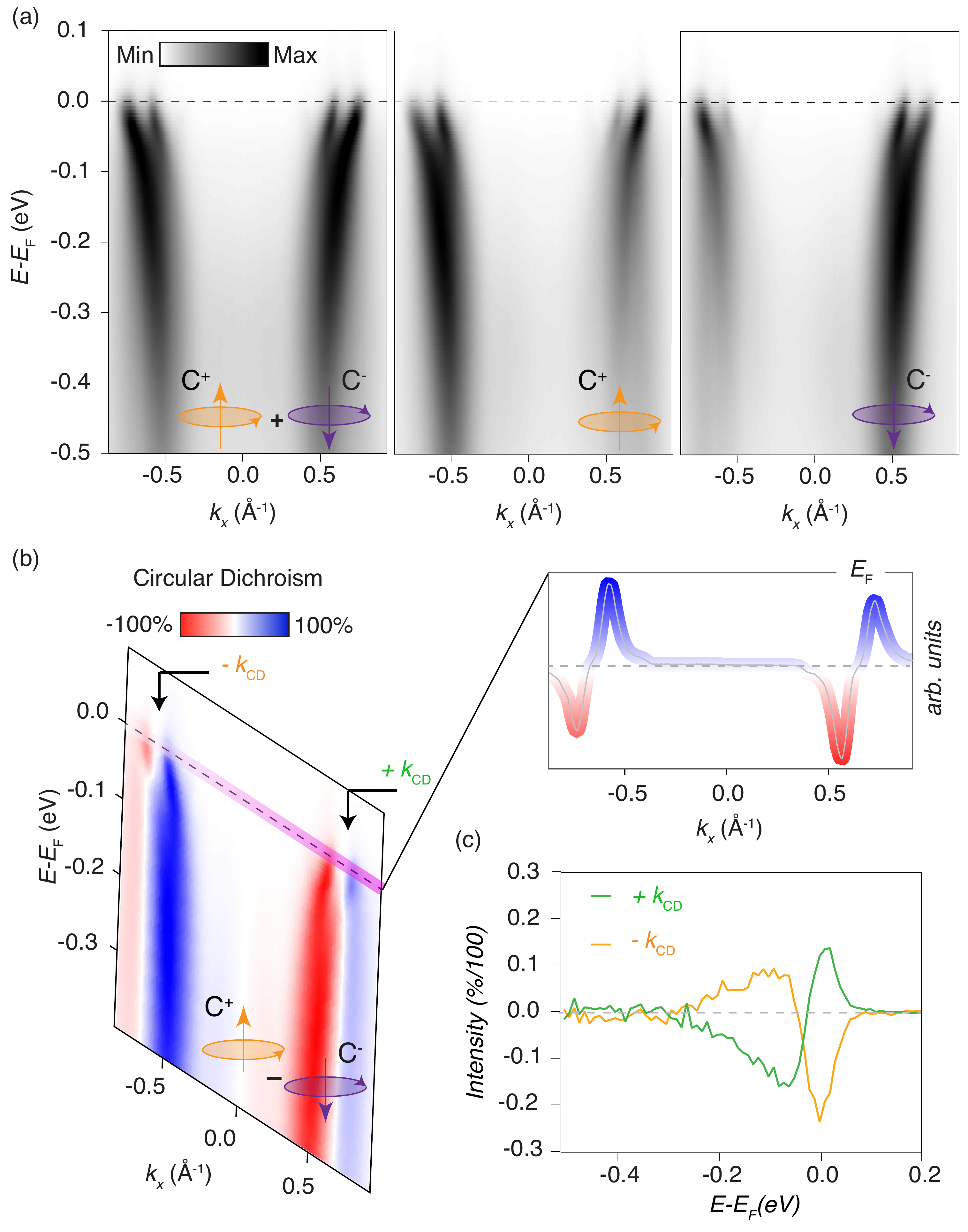}\caption{{\bf Circularly polarized spin-integrated ARPES.} {\bf a} (left) Unpolarized ARPES spectrum from Sr$_2$RuO$_4$ along the direction orthogonal to the crystal mirror plane, corresponding to the grey dashed line in Fig.\ref{fig1}c. The spectrum has been obtained by summing both contributions from (middle) right- and (right) left-circularly polarized light. Here, we will refer to as $C^{+,-}(\pm k,\uparrow,\downarrow)$ for indicating signals from right- (or left-) circularly polarized light, collected at momentum $\pm k$, and with spin-up (or down) component, respectively. {\bf b} Circular dichroism of ARPES spectrum obtained by subtracting the contributions from right and left-circularly polarized light. Remarkably, the signal changes sign from $+k$ to $-k$, with incoming light within the mirror plane. The asymmetry seen is discussed in both the main text and Methods. {\bf c} Energy-dependent circular dichroism collected with spin-detector (VLEED) at the $k$-points indicated in {\bf b}.}\label{fig2}
\end{figure*}


\begin{figure*}[!t]
\centering
\includegraphics[width=\textwidth,angle=0,clip=true]{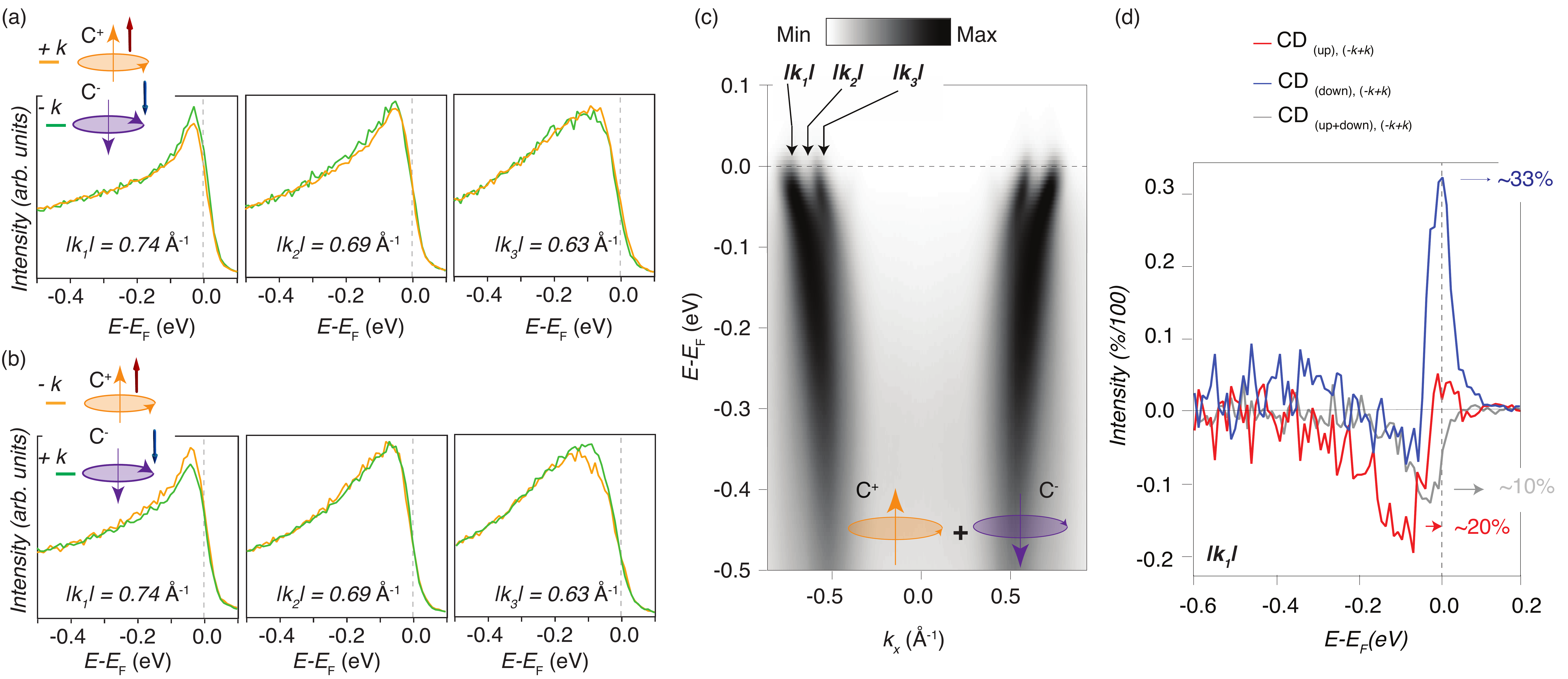}\caption{{\bf Circularly polarized spin-ARPES.} {\bf a} EDCs taken at six selected momenta ($\pm k_{i}$, $i=1,2,3$) with fixed spins and circular polarizations. In particular, the orange curves are obtained by measuring the EDCs at positive $k$ values, right-circularly polarized light and spin-up channel ($C^{+}(k,\uparrow)$), while the green curves are obtained with negative $k$ values, left-circularly polarized light and spin-down channel ($C^{+}(-k,\downarrow)$). {\bf b} ARPES spectra with reversed spin and circularly polarized light configurations: orange curves refer to $C^{+}(-k,\uparrow)$, while the green curves are obtained for $C^{-}(k,\downarrow)$. {\bf c} ARPES image indicating the $k$ values at which the EDCs have been taken. Note that both {\bf a} and {\bf b} configurations show a difference, which is larger than the experimental uncertainty. {\bf{d}} The amplitudes of the circular dichroism (at $k$-summed up to see the actual residual) are reported for both spin-integrated and spin-resolved measurements. The data show that while the spin-integrated signal (grey curve) shows a finite value, as large as 10$\%$ (which is also very similar to the experimental uncertainty of 8$\%$ as shown in \cite{DiSante_2023}), the spin-resolved channels show a significantly larger amplitude, of a factor $\times 2$ and $\times 3$ for up and down channels, respectively. The amplitude values have been extracted from the data shown in {\bf a-b} and in Methods Fig.\ref{Exp2}, after including the Sherman function and calculating the true spin polarization, as described in the Methods. The other $k$ points indicated, as well as the dichroic amplitude in MDC fashion are shown in Methods Figures \ref{Expnew}-\ref{Temp} and corroborate the validity of our result.}
\label{fig3}
\end{figure*}


In Fig.\ref{fig2}{\bf a}, we show ARPES spectra collected \FM{in the geometry of Fig.\ref{fig1}{\bf c}, compatible with signals coming from both the bulk and surface states. The latter appear weaker in spectral intensity but clearly visible with increased contrast (see Methods).} From left to right, in Fig.\ref{fig2}{\bf a}, spectra with unpolarized, right-, and left-circularly polarized photons are shown. The former shows an overall symmetric intensity pattern between features at $+k$ and at $-k$. From these spectra, the circular dichroism ($CD$) is extracted (Fig.\ref{fig2}{\bf b}), and the signal $CD(+k)$ goes to $-CD(-k)$ at opposite momenta, consistent with previous studies \cite{Nelson_2004}. This behaviour is also seen in the momentum-distribution curve (MDC) at the Fermi level (see the inset extracted along the mauve line). Importantly, an attentive look at the spin-integrated circular dichroism of  Fig.\ref{fig2}{\bf b} reveals a small asymmetry in the residual of the amplitudes (see Methods Figures \ref{Expnew}-\ref{Temp}), quantified as large as 10$\%$. This value is slightly larger than the estimated experimental error on the dichroism for this experimental setup (around 8$\%$). In addition, there might be \MC{a component} of asymmetry 
related to the character of the chiral electronic ordering (discussed in Supplementary Information). However, such an asymmetry remains significantly smaller than the one measured for spin-resolved signals. For completeness, spin-integrated data are collected with the VLEED spin detector (Fig.\ref{fig2}{\bf c} at two $\pm k$ points) and show the same behaviour.


Importantly, the spin-integrated data of Fig.\ref{fig2}{\bf c} are not only consistent with other ARPES works, but also, they do not reveal mirror-symmetry breaking relatable to an anomalous behavior of $L_z$. On the other hand, quantities such as \MC{spin-orbital chiral} currents, which strongly depend on the spin as much as on the orbital angular-momentum, cannot be imaged by standard circularly polarized ARPES. Here, we will refer to $C^{+,-}(\pm k,\uparrow,\downarrow)$ for indicating signals from right- (or left-) circularly polarized light, collected at momentum $\pm k$, and with a spin-up (or down) component, respectively. In a perfectly symmetry-preserving situation, the circularly-polarized spin-ARPES intensity transforms under the mirror operator, from $C^{+}(+k,\uparrow)$ to  $C^{-}(-k,\downarrow)$ (or equivalently from $C^{+}(-k,\uparrow)$ to  $C^{-}(+k,\downarrow)$). This means that $C^{+}(+k,\uparrow)$ and $C^{-}(-k,\downarrow)$ (or $C^{+}(-k,\uparrow)$ and $C^{-}(+k,\downarrow)$) are expected to be the same under mirror symmetry in the case that the latter is preserved \cite{Schuler_2020}. However, if loop currents are present along the surface, the mirror symmetry is broken and these quantities are not equivalent anymore. This scenario \FM{is what we tested with CP-Spin-ARPES and reported in Fig.\ref{fig3}{\bf a-b}}: \FM{we observe subtle differences at different $k$ points, as noted in Fig.\ref{fig3}{\bf c}, between $C^{+}(+k,\uparrow)$ and $C^{-}(-k,\downarrow)$ (or $C^{+}(-k,\uparrow)$ and $C^{-}(+k,\downarrow)$). These differences, despite {being} very small, result instead {in} a sizeable asymmetry in the amplitudes of spin-up and down dichroism (Fig.\ref{fig3}{\bf d}). Note that for the latter, positive and negative momenta have been summed, compensating for possible instrumental asymmetry of the measurements.}

\FM{Such a mirror-symmetry breaking in the amplitude of the spin-dichroism seems compatible with the presence of spin-orbital quadrupole currents, as predicted by the theory.} In addition to this, the finite dichroism difference observed experimentally is not seen for a temperature of 77~K, higher than the magnetic transition temperature as indicated by the muons spectroscopy \cite{Fittipaldi_2021}, however, this requires further investigation as the thermal broadening becomes significantly more pronounced (see Methods). \FM{The important finding here, is that the estimated difference in spin from the dichroic signals is up to a factor {of 3 times} larger than the spin-integrated one {.} In Fig.\ref{fig3}{\bf d}, the amplitude of the spin-integrated dichroic signal is approximately 10$\%$ (grey curve), while the spin-resolved is as high as {20$\%$ for spin-up (red curve) and 30$\%$  for spin-down (blue curve);} see Supplementary Information \MC{for possible explanation} of the spin-integrated asymmetry observed. We emphasize that a quantitative analysis is difficult and the signals detected are subtle. To quantify these effects correctly, future measurements as a function of photon energy and various geometries will be desirable. Nevertheless, the presence of a sizeable asymmetry in the amplitude signal from $k$ and $-k$ is observed and this is consistent with the theoretical predictions.}
\\

\begin{figure*}[!t]
\centering
\includegraphics[width=\textwidth,angle=0,clip=true]{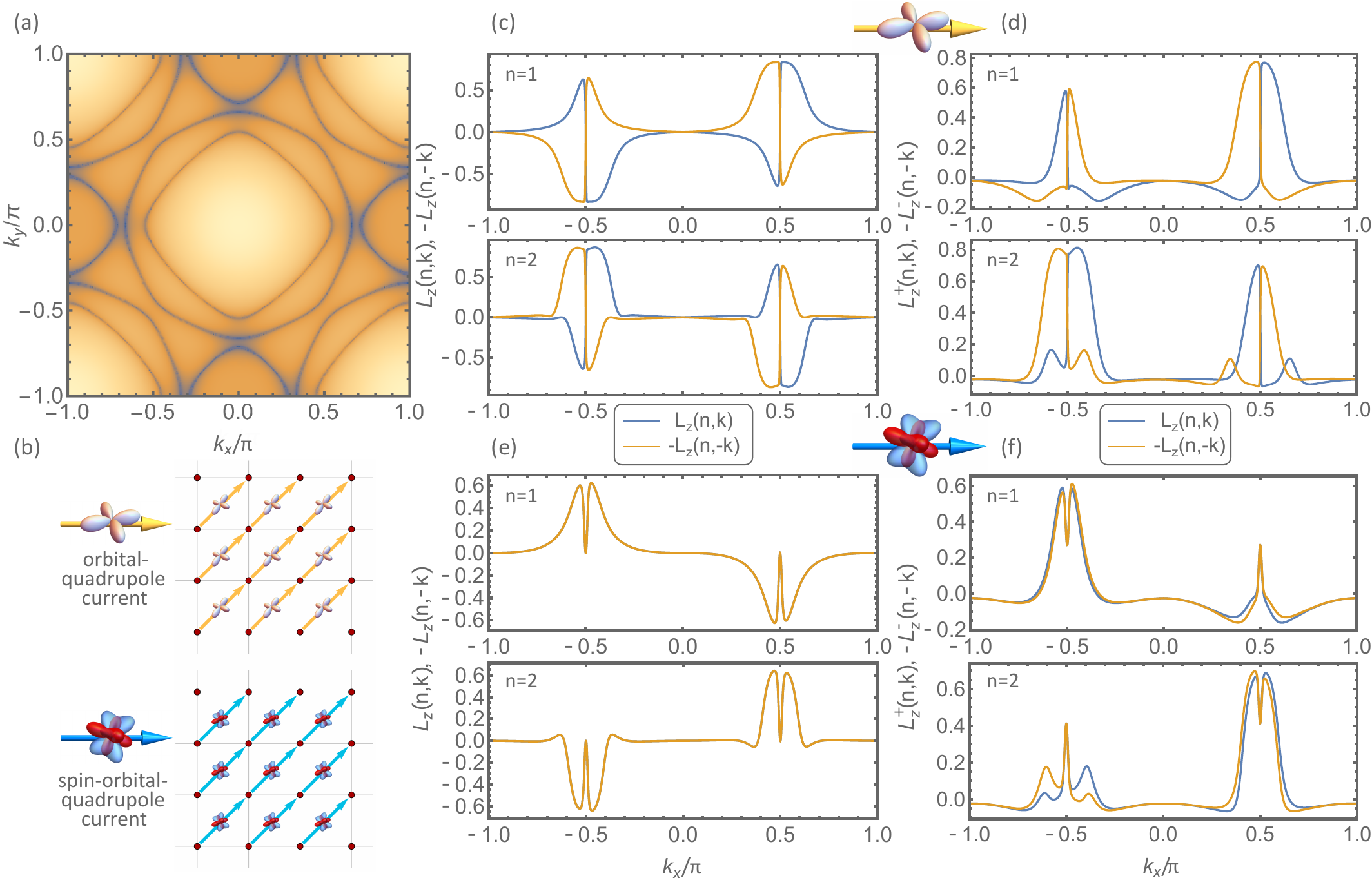}\caption{{\bf Orbital and spin-orbital textures in the presence of chiral currents.} {{\bf a} Computed Fermi surface of the Sr$_2$RuO$_4$. 
{\bf b} Broken symmetry state with an electronic pattern marked by either orbital-quadrupole (top panel) or spin-orbital quadrupole currents (bottom panel), The sketch indicates a current in real space connecting the Ru sites along the [110] direction. {\bf c} Electronic phase with chiral orbital-quadrupole currents: amplitude of the orbital angular momentum ${L_z(n,k)}$ of the bands $|\psi_{n,k}\rangle$ with $n=1,2$ evaluated along the ${\Gamma-X}$ direction (${L_z(n,k)}=\langle \psi_{n,k}| \hat{L}_z |\psi_{n,k}\rangle$). For clarity, we plot both ${L_z(n,k)}$ (blue) and $-{L_z(n,-k)}$ (yellow) for any given momentum ${k}$ to directly compare the amplitudes at opposite momenta. {\bf d} Electronic phase with chiral orbital-quadrupole currents: amplitude of the spin projected orbital angular momentum related to the out-of-plane spin-up ($+$) and down ($-$) components as selected by the projector $(1\pm \hat{s}_z)$. The amplitude is given by ${L^{\pm}_z(n,k)}=\langle \psi_{n,k}| (1\pm \hat{s}_z) \hat{L}_z |\psi_{n,k}\rangle$. The amplitudes of ${L_z(n,k)}$ and ${L^{\pm}_z(n,k)}$, displayed in {\bf c} and {\bf d}, do not show any symmetry and do not match at ${k}$ and $-{k}$. {\bf e} Electronic phase with chiral spin-orbital quadrupole currents with antisymmetric $L\,S$ content with respect to the current flow direction $k$ (i.e. $ {\bf k} \cdot (\hat{\bf{L}}\times \hat{{\bf{s}}})$:  for this configuration ${L_z(n,k)}$ and ${L_z(n,-k)}$ coincide. {\bf f} Electronic phase with chiral spin-orbital quadrupole currents with antisymmetric $L\,S$ combination: spin projected orbital moment ${L^{\pm}_z(n,k)}$ at opposite momenta are unequal in amplitude. Similar trends occur for the other bands (see Supplementary Information).}}
\label{fig4}
\end{figure*}

\noindent{\bf{\large Chiral currents phase}}\\
We now analyze our experimental results from a theoretical point of view. 
This can be done with a 
two-dimensional tight-binding description of the electronic structure of Sr$_2$RuO$_4$ based on the Ru $d$-orbitals ($d_{xy}$,$d_{xz}$,$d_{yz}$) in such a way to capture the profile of the experimental Fermi surface (Fig.4{\bf a}). In this model, we include the broken symmetry states hosting orbital and spin-orbital quadrupole currents that are driven by the $d-d$ Coulomb interactions (see Supplementary Information for details). 
The internal structure of the charge currents is provided by either the orbital quadrupole $\hat{L}_p \hat{L}_q$ or by the spin-orbital quadrupole $\hat{L}_p \hat{s}_q$ tensors (see also Fig.\ref{fig1}{\bf{a}} for differences between these currents). For a given direction $k_l$ in momentum space, the amplitude of the charge current propagating through the lattice can in principle contain various components of the type $\hat{j}^{l}_o=\sin(k_l) \hat{L}_p \hat{L}_q$ and $\hat{j}^{l}_{so}= \sin(k_l) \hat{L}_p \hat{s}_q$ for the orbital and spin-orbital quadrupole, respectively (note that the first does not contain any spin channel, the second instead does). 
In the presence of currents flowing along e.g. $l=x$ that break all mirror symmetries (i.e. $M_{j=x,y,z}$), the orbital and spin-orbital quadrupoles have to include components with $\hat{s}$ and $\hat{L}$ that are perpendicular to $x$, thus lying in the $yz$ planes. For instance, a term of the type $\sin(k_{x}) \hat{L}_{y} \hat{L}_{z}$ does not preserve the mirror symmetry for any choice of the $M_{j=x,y,z}$ transformations. The latter is exactly the case of Sr$_2$RuO$_4$, where a uniform charge current flows with the direction of the momentum aligned along the $l=[110]$ direction of the Ru lattice (see Fig. \ref{fig4}{\bf b}). This pattern is compatible with symmetry-allowed loop currents involving charge current flowing from ruthenium to oxygen atoms at the octahedra length scale \cite{Fittipaldi_2021}. 
\MC{The qualitative outcomes of the results are not altered by surface reconstruction or by having currents flowing along other symmetry directions (see Supplementary Information for details).}
We select a spin-orbital chiral state that breaks the $C_4$ rotational symmetry \MC{because it is compatible with the scanning tunnelling microscopy findings \cite{Marques_2021}.} 
However, spin-orbital chiral states with loop currents that are rotational invariant can be also constructed. They break translational symmetry and do not modify the qualitative outcomes of the analysis (see Supplementary Information).

Additionally, the charge current along [110] is directly relevant when probing the electronic states along the corresponding direction of the Brillouin zone (along $\Gamma$-X as in our experiment). To evaluate the orbital and spin-orbital textures for the electronic states in the presence of either orbital or spin-orbital quadrupole currents, we will focus on such direction. For each band eigenstate $|\psi_{n,k}\rangle$ at {a given momentum $k$,} we determine the amplitude of the out-of-plane orbital moment ${L_z(n,k)}=\langle \psi_{n,k}| \hat{L}_z |\psi_{n,k}\rangle$ {and} the spin-projected orbital moment ${L^{\pm}_z(n,k)}=\langle \psi_{n,k}| (1\pm \hat{s}_z) \hat{L}_z |\psi_{n,k}\rangle$ {(i.e. the out-of-plane spin-up ($+$) and down ($+$) components as singled out by the projector $(1\pm \hat{s}_z)$)}. These observables are related to the dichroic and spin-dichroic amplitudes probed by ARPES, respectively (see Supplementary Information for details). For clarity and simplicity, {we} display only two representative bands, e.g. $n=1,2$ (the behavior for the other bands is however qualitatively similar and reported in Supplementary Information). As shown in Fig.\ref{fig4}{\bf c}, a broken symmetry state with orbital quadrupole currents exhibits an asymmetry in the orbital angular moment at opposite momenta, i.e. ${L_z(n,k)}\neq -{L_z(n,-k)}$. Moreover, both the amplitude and the sign of ${L^{\pm}_z(n,k)}$ and ${L^{\pm}_z(n,-k)}$ are dissimilar (Fig.\ref{fig4}{\bf d}). Instead, for spin-orbital currents (Fig.\ref{fig4}{\bf b}) with an antisymmetric combination of $L$ and $S$, i.e. $j^{l}_{so}=\sin(k_l) (\hat{L}\times \hat{s})_l$, we find that the orbital angular momentum turns out to be \MC{antisymmetric} at $k$ and $-k$, namely ${L_z(n,k)}=-{L_z(n,-k)}$ (Fig.\ref{fig4}{\bf e}), while the spin-projected orbital moment does not exhibit any symmetry relation among the states at opposite momentum (Fig.\ref{fig4}{\bf f}). Importantly, the last asymmetry is the same observed experimentally by the CP-Spin-ARPES. One way to grasp the origin of this behavior is to inspect the structure of the equations of motion for the amplitudes of the orbital ${L_z(n,k)}$ and spin ${s_z(n,k)}$ moments \MC{(see Supplementary Information for details)}. \MC{The chiral currents lead to spin-orbital torques that in the case of the spin-orbital quadrupole current result into a balance such as the amplitude of the orbital moment has a symmetric behavior at symmetry-related momenta.}

Additionally, we have also addressed the role of sublattice chiral currents due to the surface reconstruction resulting from the octahedral rotations (see Supplementary Information). 
\MC{When we consider an inhomogeneous state with a staggered amplitude modulation of the currents, we find that a small asymmetry of the orbital moment is obtained. Nevertheless, its amplitude is substantially smaller than that of the spin-projected orbital moment (see Supplementary Information). This implies that non-homogeneous chiral currents are also compatible with the observation of a much larger spin-dichroic asymmetry as compared to the dichroic one.}
\MC{This behavior holds independently of the selected band and momentum, differently from that of magnetic states, e.g. antiferromagnetic, with a pattern in the spin and orbital moments that break time and mirror symmetries (see Supplemental Information)}.
\\

\noindent{\bf{\large Conclusions}}\\
In conclusion, we discovered a spin-orbital and angular-momentum sensitive methodology based on CP-Spin-ARPES, able to probe symmetry breaking compatible with the existence of spin-orbital chiral currents. \MC{While the study has been applied to the archetypal Sr$_2$RuO$_4$}, this methodology is general to all chiral surface metals and constitutes a tantalizing experimental avenue to detect symmetry-broken chiral states. 
\MC{However, since the effect is subtle one cannot exclude that the observed symmetry breaking might arise from other real space chiral orderings that break both time and mirror symmetries. Nevertheless, our work stimulates the combined use of circular dichroism and spin-selective photoemission to investigate how all three quantities $L$, $S$, and $LS$ behave and their relationship to the crystal symmetries, which are markers for hidden ordered phases}. 

The spin-dichroic signal we have used to detect the putative presence of quadrupole spin-orbital currents at the surface of Sr$_2$RuO$_4$ can be used without restrictions in other quantum materials, even when currents appear in the bulk of a (centrosymmetric) crystal. In this situation, the currents might preserve the combination of time-reversal with inversion symmetry, with the consequent absence of dichroism. Nevertheless, the asymmetry of the spin-dichroic signal could be still visible and therefore represents an efficient diagnostic tool for spin-orbital chiral metallic phases.
\\

%

\newpage

\noindent{\bf{\large Methods}}\\
The samples of Sr$_2$RuO$_4$ were grown by floating zone technique, following the procedure of Ref.\cite{Fittipaldi_2005_M}. Single crystals were post-cleaved in ultrahigh vacuum at a base pressure of 1$\times 10^{-10}$~mbar and at a temperature of $20$~K (and $77$~K). The latter was kept constant throughout the measurements. The experiment was performed at the NFFA-APE Low Energy beamline laboratory at the Elettra synchrotron radiation facility designed with an APPLE-II aperiodic source for polarized EUV radiation and a vectorial twin-VLEED spin polarization detector downstream of a DA30 Scienta ARPES analyser \cite{Bigi_2021_M}. The photon energy used for our measurements was $40$~eV, which is found to maximize the spectral intensity, as also shown in Ref.\cite{Sunko_2019_M}. The energy and momentum resolutions were better than $12$~meV and $0.018$~\AA$^{-1}$, respectively. Importantly, as already mentioned, we stress that in order to eliminate the geometrical contribution to the circular polarization, the crystals were aligned as in Fig.\ref{fig1}{\bf c-d}. For completeness, we report seminal works on ARPES and dichroism which might help the understanding of the measurements performed in our article, i.e., Ref.\cite{Beaulieu_2020_M, Scholz_2013_M, Park2012_M, Cho_2018_M, DiSante_2023_M}.

In this part of the Methods, we report additional measurements which contributes to corroborate the message and conclusions in the main text.

\begin{figure*}[h!]
\centering
\includegraphics[width=0.4\textwidth,angle=0,clip=true]{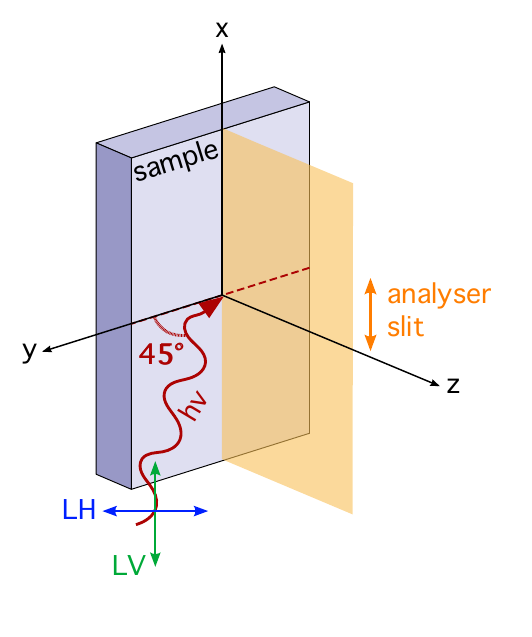}\caption{{\bf Photoemission experimental geometry.} The sample, represented by the purple box, is such that the incoming synchrotron radiation (red wavy arrow) impinges with an angle of 45$^\circ$ with respect to its surface. In this configuration, with linear polarizations, we would have linear vertical (LV, green double-headed arrow) lying completely on the sample surface. Instead, linear horizontal (LH, blue double-headed arrow) would have both in- and out-of-plane components, projected along the $y$- and $z$-axis, respectively. The slit of the analyser is along the scattering plane (vertical slit).}\label{M1}
\end{figure*}

\begin{figure*}[h!]
\centering
\includegraphics[width=0.7\textwidth,angle=0,clip=true]{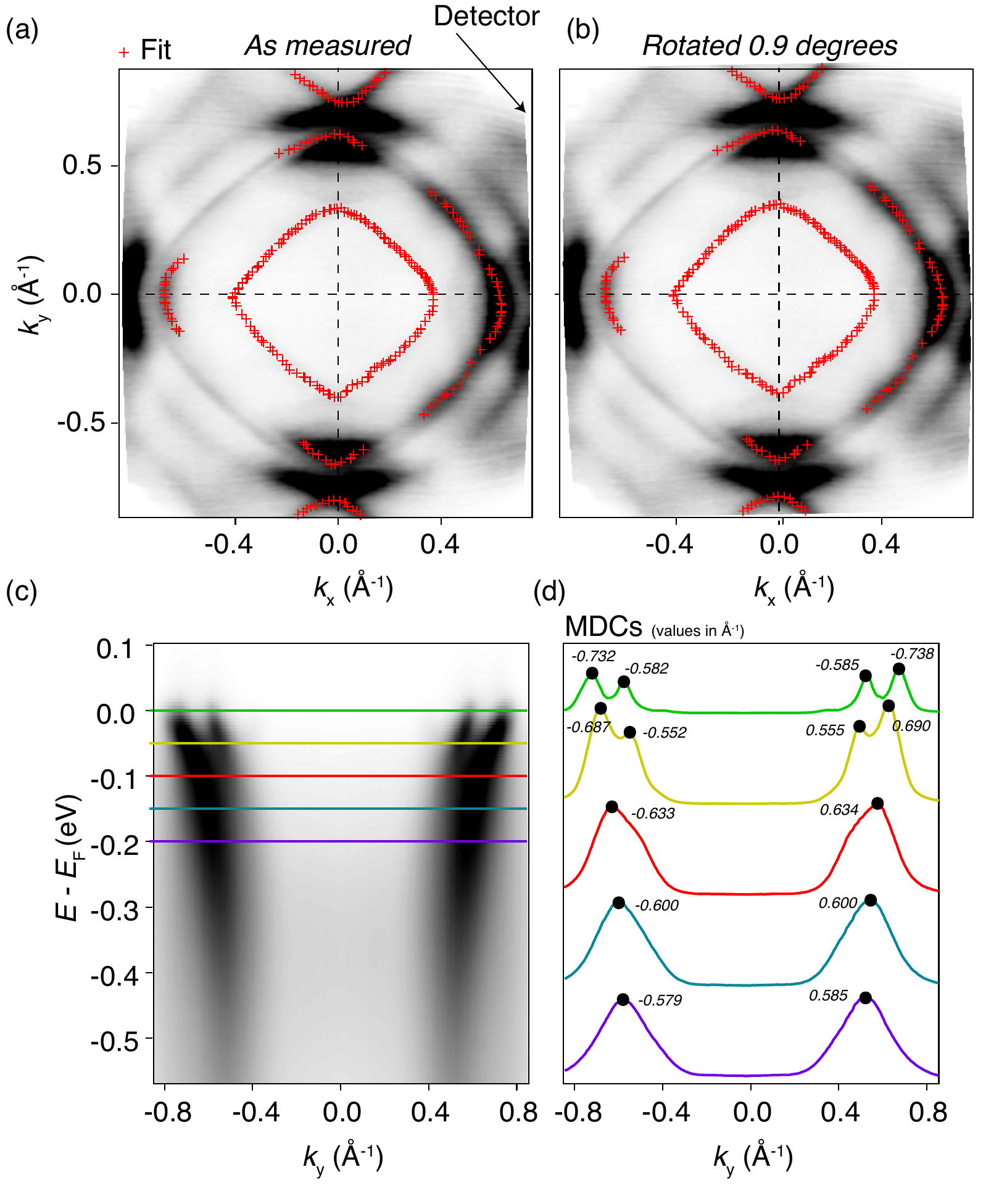}\caption{{\bf ARPES identification of the surface states and sample alignment.} {\bf a} Fermi surface collected at 40 eV (sum of the two circularly polarized lights) showing both bulk bands and surface states. The latter are weaker than the bulk in intensity but still visible. To better appreciate the precise sample alignment we fitted the data and extracted the $k$ positions, reported in the image as red markers. The mirror plane deviates from the ({\bf b}) ideal condition by 0.9$^\circ$. {\bf c}) Energy versus momentum dispersion collected in the same experimental conditions of (Fig.\ref{M1}{\bf a)}) showing a very symmetric character. To better appreciate this, we extracted MDCs and plotted them in panel {\bf d} along with their extracted $k$ values.}\label{M2}
\end{figure*}

\noindent {\bf{Sample alignment and experimental geometry}}\\
\FM{When using circularly polarized light, the disentanglement between geometrical and intrinsic matrix elements is crucial and challenging. A solution is to have the incoming radiation exactly within one of the mirror planes of the system studied and to measure in the direction orthogonal to that plane, as we show in Fig. 1c of the main text. In such a configuration, the differences in the CP-Spin-ARPES signal can be attributed to intrinsic differences in $LS$ and the geometrical contributions are well-defined. In this regard, it is of paramount importance to align the sample very carefully. In the present case, the symmetric character of the material's Fermi surface \cite{Damascelli_2000_M, Tamai_2019_M, Sunko_2019_M} allows us to carefully align the sample with the incoming beam of photons lying within a mirror plane. The alignment of the sample was carried out by monitoring the experimental Fermi surface and by making sure that the analyzer slit direction was perpendicular to the mirror plane. As one can see in Figs. \ref{M1} and \ref{M2}, we estimated our alignment to be better than 0.9$^\circ$ from the ideal configuration, a value within the uncertainty considering our angular azimuthal precision ($\approx$1$^\circ$). Furthermore, different samples mounted gave us the same results, corroborating the robustness of the measurements outputs within this azimuth uncertainty.}

In the laboratory APE-LE, our sample is placed in the manipulator in normal emission conditions, with the synchrotron light impinging on the sample surface at 45$^\circ$ angle. This means that standard linear polarizations, such as linear vertical and linear horizontal (See Fig. \ref{M1}), would act quite differently on the matrix elements selection rules: In particular, linear vertical light would be fully within the sample plane, while linear horizontal would have a component within the plane and one out-of-plane (with 50$\%$ intensity each). Now, when using circularly polarized light, in order to distinguish between real and geometrical matrix element effects, the incoming light needs to be aligned within the experimental error, within one of the mirror planes of the sample.

To estimate the azimuthal value we have performed fitting of the $k$-loci of the Fermi surface contours (red markers in Fig.\ref{M2} a-b) and we have then aligned the horizontal and vertical axis (see the next subsection for fitting details). As one can see, in our configuration, there is negligible misalignment between the states at positive and negative $k$, see Fig. \ref{M2} c-d for example. The latter shows, that by extracting momentum-distribution curves (MDCs), coloured horizontal lines in Fig. \ref{M2} c) the peak positions are symmetric within the resolutions of the instrument (~12 meV for energy and $0.018$~\AA$^{-1}$). Therefore, we are in a good condition to safely perform the measurements shown in the main text.

\noindent{\bf{Details of the fitting}}\\
The Fermi surfaces $k$-loci shown in Fig. \ref{M2} a-b and the positions of the peaks in Fig.\ref{M1} d have been extracted by fitting the ARPES data. The fitting procedure used is standard and consists of fitting both EDCs and momentum-distribution curves by using Lorentzian curves convoluted by a Gaussian contribution which accounts for the experimental resolutions. Then, as part of the fit results, we extracted the $k$ positions of the peaks, which are those reported as red markers in Fig. \ref{M2} and the values in Fig. \ref{M2} d.

\noindent{\bf{Spin-ARPES data}}\\
For the extraction of the values reported, the spin-data shown in the manuscript have been also normalized to include the action of the Sherman function of the instrument. In particular, the data for spin-up and spin-down channels have been normalised to their background, such that this matched in both cases. In the present study, the background normalisation was executed on the high-energy tails of the EDCs far from the region where the actual spin polarization is observed. After normalisation, in order to extract the spin intensity, we used the following relationships:\\

\begin{center}
$I^{TRUE}(k,\uparrow)=\frac{I^{TOT}(k)}{2}*(1+P)$
\end{center}

\begin{center}
$I^{TRUE}(k,\downarrow)=\frac{I^{TOT}(k)}{2}*(1-P)$
\end{center}

where $P$ is the polarization of the system and $I^{TOT}=I^{bg.norm}(k,\uparrow)+I^{bg.norm}(k,\downarrow)$ is simply the sum of the intensity for EDCs with spin-up and spin-down after normalisation to the background. For the polarisation $P$, the Sherman function from the instrument was included (i.e. $\eta$=0.3 \cite{Bigi_2021_M} and calibrated from measurements on gold single crystal). This is described by:\\

\begin{center}
$P(k)=\frac{1}{\eta}*\frac{I^{bg.norm}(k,\uparrow)-I^{bg.norm}(k,\downarrow)}{I^{bg.norm}(k,\uparrow)+I^{bg.norm}(k,\downarrow)}$
\end{center}

\begin{figure*}
\centering
\includegraphics[width=0.8\textwidth,angle=0,clip=true]{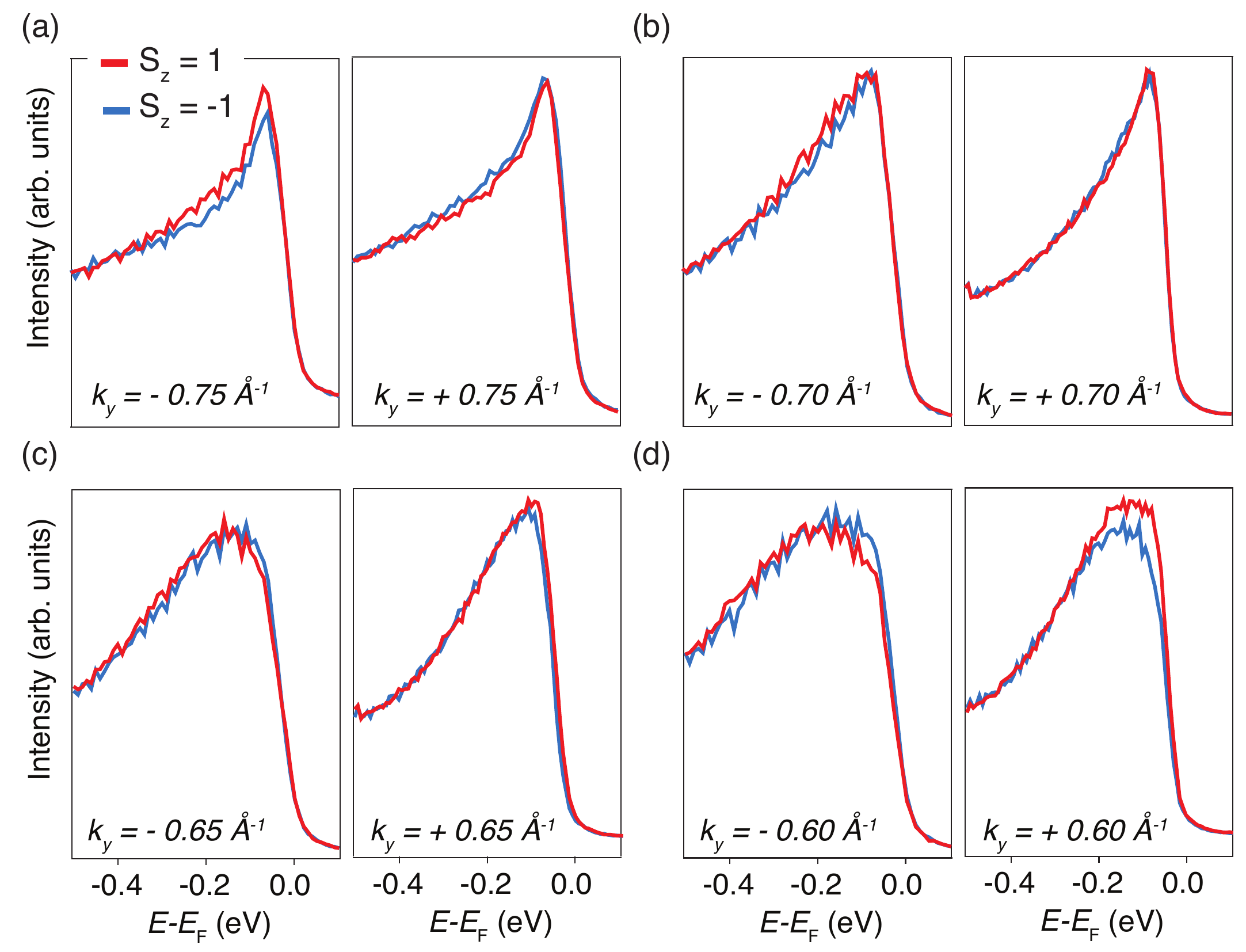}\caption{{\bf Spin-resolved data with unpolarised light.} Energy distribution curves collected at momenta {\bf a-d} $k_y=0.75$ \AA$^{-1}$, $k_y=0.70$ \AA$^{-1}$, $k_y=0.65$ \AA$^{-1}$, $k_y=0.60$ \AA$^{-1}$. The data are with sum of circular right and left light and spin-up and spin-down channels have been shown in red and blue, respectively.}\label{Exp2}
\end{figure*}

This procedure was done for all light polarizations. Additionally, we also characterized the spin channels by using different polarization vector directions, as shown in Fig.\ref{Exp2}.

\noindent{\bf{Dichroism and spin-dichroism amplitudes}}\\
A way to visualise the breaking of the time-reversal symmetry is to analyse the dichroic signal already shown in Fig. 2c of the main text but resolved in the two different spin channels, i.e. up and down, which gives rise to very different amplitude values when measured at $\pm k$ (expected for time-reversal symmetry breaking but not expected otherwise). \FM{We show this here at a selected momentum values. The amplitude values have been extracted from the data shown in {\bf a-b} and in Fig. \ref{Exp2}, after including the Sherman function normalization.}

\begin{figure*}
\centering
\includegraphics[width=\textwidth,angle=0,clip=true]{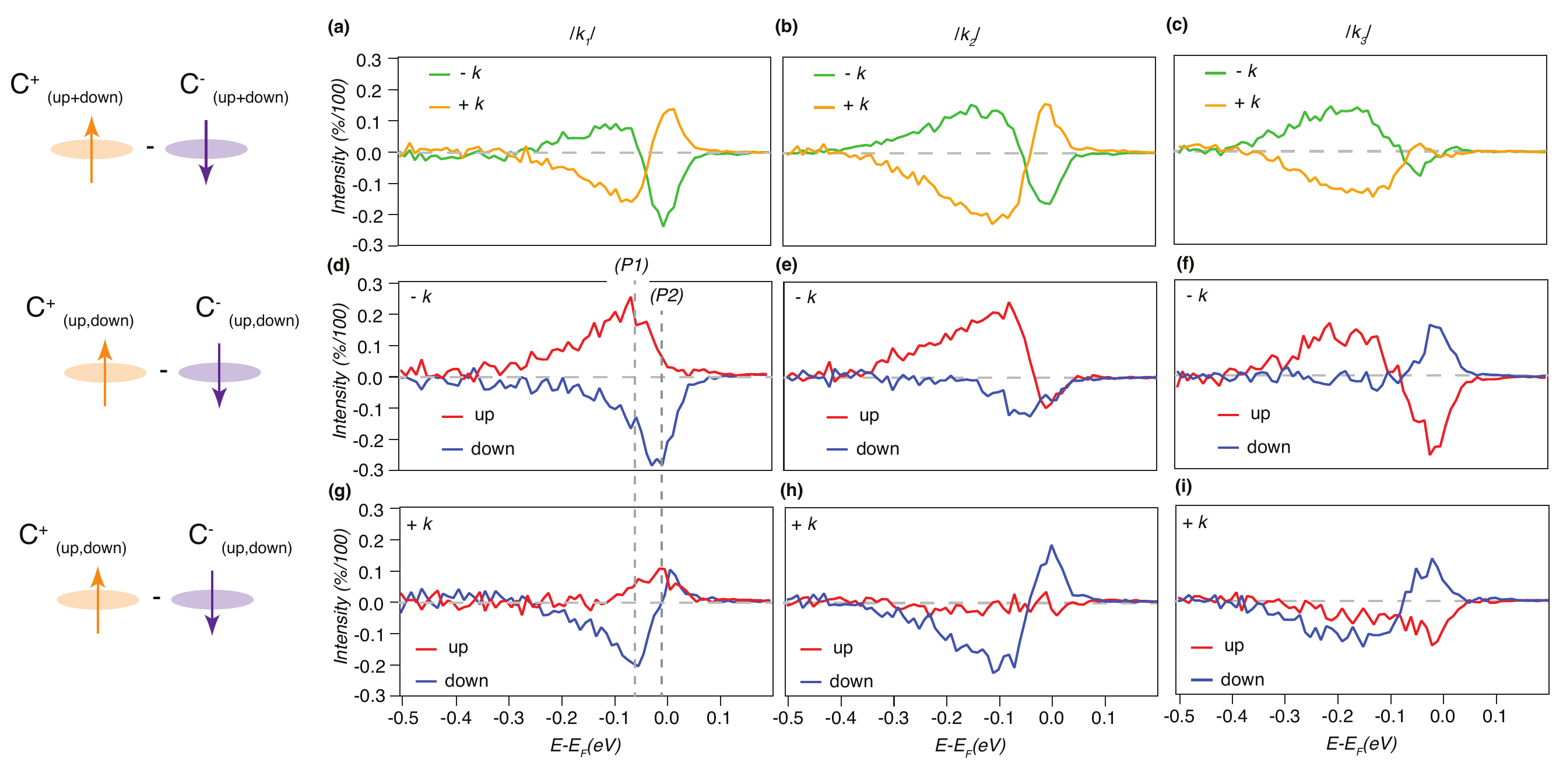}\caption{\FM{{\bf Spin-integrated and spin-resolved dichroism.} Spin integrated circular dichroism collected at {\bf{a}} $\bm{k_y}=\pm{0.73}{\AA}^{-1}$ ($k_1$), {\bf{b}} $\bm{k_y}=\pm{0.68}{\AA}^{-1}$ ($k_2$), and {\bf{c}} $\bm{k_y}=\pm{0.72}{\AA}^{-1}$ ($k_3$), as indicated in the main text Figure 3c. Green curves indicate negative $k$, and orange curve positive $k$. {\bf{d-e-f}} Spin-resolved circular dichroism collected at negative $k$ for the three momenta indicated. {\bf{g-i}} Same but collected at positive momenta.
}}\label{Exp3}
\end{figure*}

\begin{figure*}
\centering
\includegraphics[width=\textwidth,angle=0,clip=true]{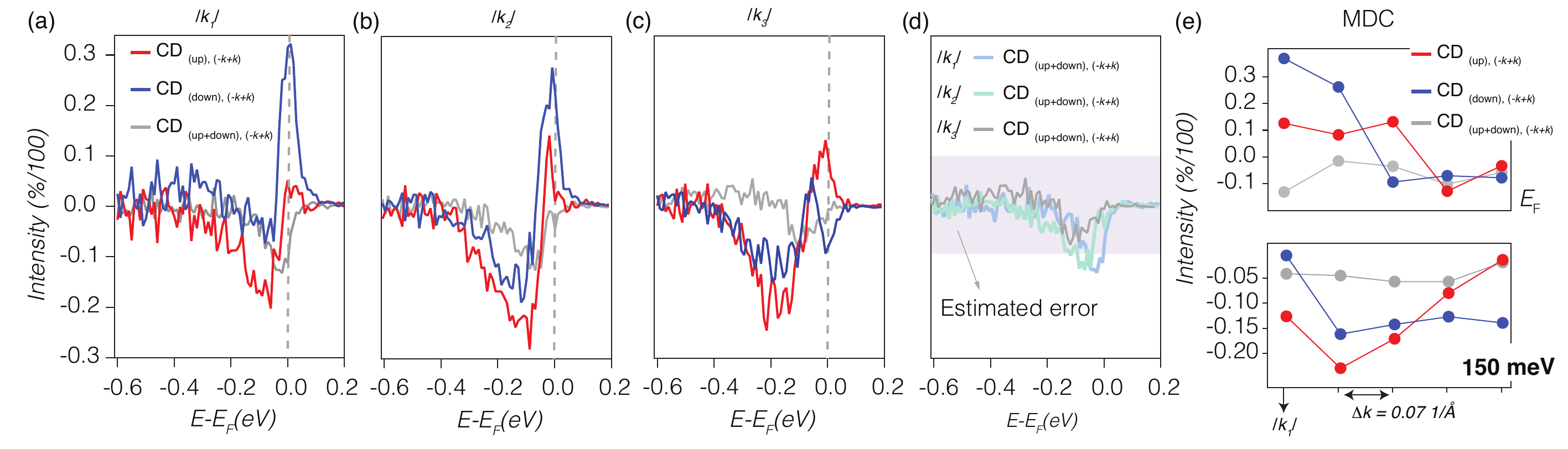}\caption{\FM{{\bf Amplitude of the dichroism, EDC and MDC.} {\bf{a-b-c}} The amplitudes of the dichroism (at $k$-summed up to see the actual residual) are reported for $k_{1,2,3}$. These show that, while {\bf d} the spin-integrated signal (grey curve) shows a finite value, as large as 10$\%$ (which is also very similar to the experimental uncertainty as reported in \cite{DiSante_2023_M} - purple stripe), the spin-resolved channels indicate a significantly larger amplitude, of a factor larger than $\times 2$ and $\times 3$ for up and down channels, respectively. {\bf{e}} The amplitude of the dichroism have been also collected by using MDC at two binding energies, i.e., at the Fermi level and at 150 meV below it. As one can see, the grey line, which is the spin-integrated dichroism is nearly flat (in average is 7$\%$ - obtained by summing up all the points), while the spin-up and spin-down channels are varying and well-different. The fact that these are also varying is quite remarkable and indicates that our signal is intrinsic in nature.}}\label{Expnew}
\end{figure*}

To better understand the claim in the main text, we here show in Fig.\ref{Exp3} the relative amplitudes of the dichroic versus spin-dichroic signal. First of all, let us focus on the spin-integrated dichroism shown in Fig.\ref{Exp3}a. Here, the orange and green curves represent positive and negative $k$ values, respectively and their behaviour is overall symmetric with respect to zero. However, a small asymmetry can still be noticed, estimated as large as ~10$\%$, a value which is very close to previously reported ones \cite{DiSante_2023_M} (~8$\%$). As we will clarify from a theoretical point of view, a small degree of asymmetry in the spin-integrated dichroism can be still expected, however, the amplitudes of the dichroism selected in their spin channels are supposed to be quite larger. To demonstrate this difference, we have shown how the dichroism curves, resolved in their spin channels (up-red and down-blue), appear at negative $k$ (Fig.\ref{Exp3}d-f) and at positive $k$ (Fig.\ref{Exp3}g-i). By also considering their residuals, we can compare them to the amplitude of the spin-integrated signal. We reported this comparison in Fig.\ref{Expnew}: The spin-down channel shows an amplitude as high as 30$\%$, the spin-up as high as 20$\%$. These values are three times and two times bigger, respectively, than the residual extracted for the spin-integrated signal. Such a large difference corroborates the validity of our methodology and the claims of our work. Note that by summing the positive and negative momentum is also counteracting possible effects due to small sample misalignment.

\noindent{\bf{Data and temperature}}\\
For completeness, we have also performed $C^{+}(+k, \uparrow)$ and $C^{-}(k, \downarrow)$ on the sample after cleaving it also at high temperature (HT, 70~K), which is above the magnetic transition of Sr$_2$RuO$_4$. We report the results in Fig.\ref{Temp}. In particular, in  Fig.\ref{Temp} a-c, the upper panels with blue lines show the difference between $C^{+}(+k, \uparrow)$ and $C^{-}(k, \downarrow)$ - normalized by their sum - at three values of $k$ and at low temperature (LT), while the lower line is the same for the data collected at 70 K. If in the LT configuration we do observe a varying finite signal, at HT we did not see such a variation. It is important to mention that even with our resolution, we do not see any finite signal, there might be still some differences which could be observed above the magnetic transition, because it is likely that not all magnetic excitations are turned off immediately, however, a reduction should be still observed. In addition, the HT data are more noisy: even if we cleaved the samples at HT, and as the ARPES shown in Fig.\ref{Temp} d-e confirms their presence, they are much weaker than in LT and broadened thermally. Such a thermal broadening is not surprising to see in ARPES. Nevertheless, even if with reduced intensity, the surface states are still well visible.

\begin{figure*}
\centering
\includegraphics[width=0.8\textwidth,angle=0,clip=true]{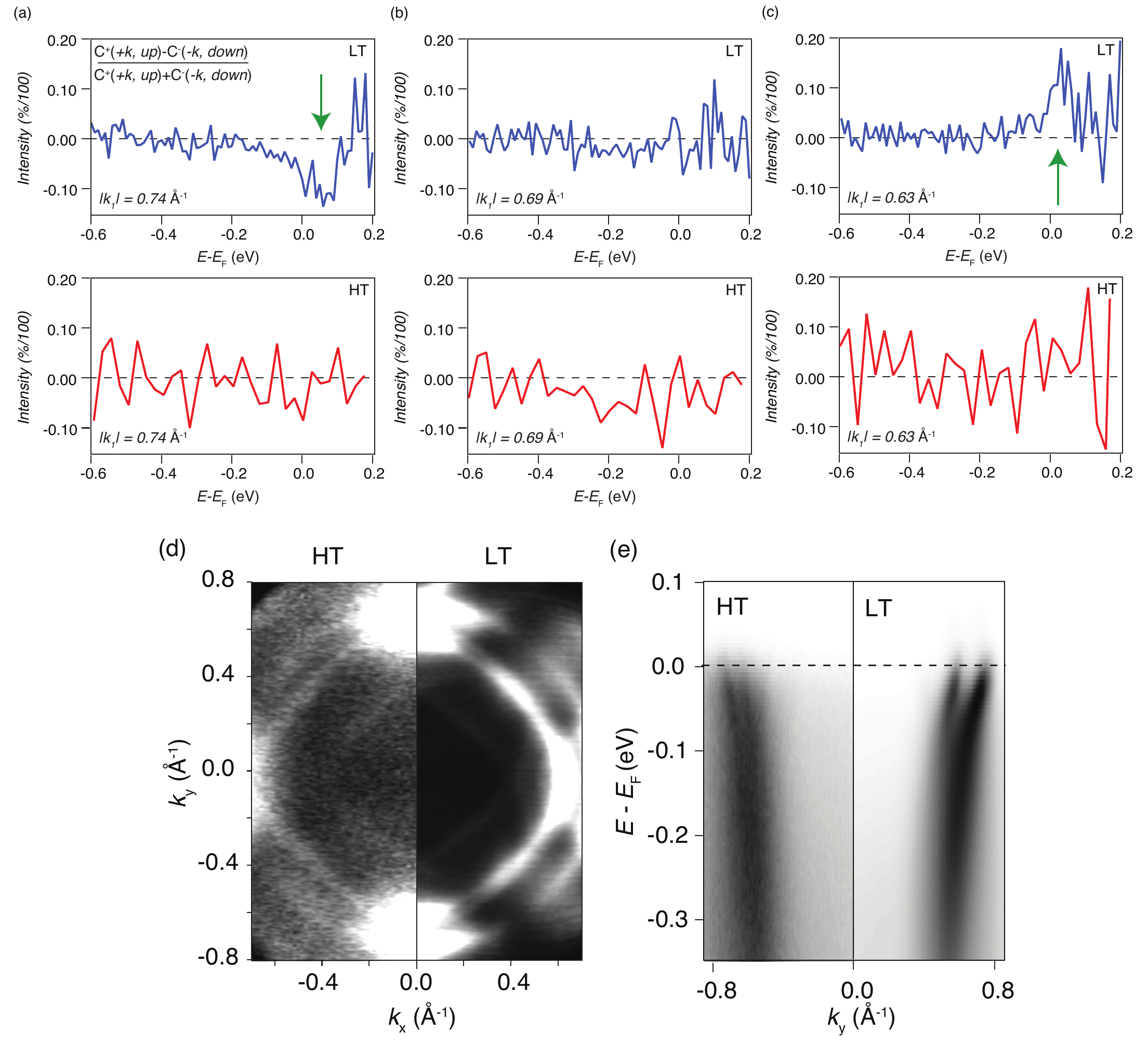}\caption{{\bf Temperature dependence data.} {\bf a-c} The upper line (blue curves) shows the difference between $C^{+}(+k, \uparrow)$ and $C^{-}(k, \downarrow)$ - normalized by their sum - at three values of $k$ and at low temperature (LT), while the lower line is the same for the data collected at 70 K (high temperature, HT), above the magnetic transition. When at LT we do observe a varying finite signal, it starts from negative and it switches sign into positive as a function of $k$, at HT we did not see such a variation. {\bf d} and {\bf e} show the Fermi surface maps and energy versus momentum dispersion for both LT and HT data. The surface states are visible in both cases.}\label{Temp}
\end{figure*}

\begin{figure*}
\centering
\includegraphics[width=0.5\textwidth,angle=0,clip=true]{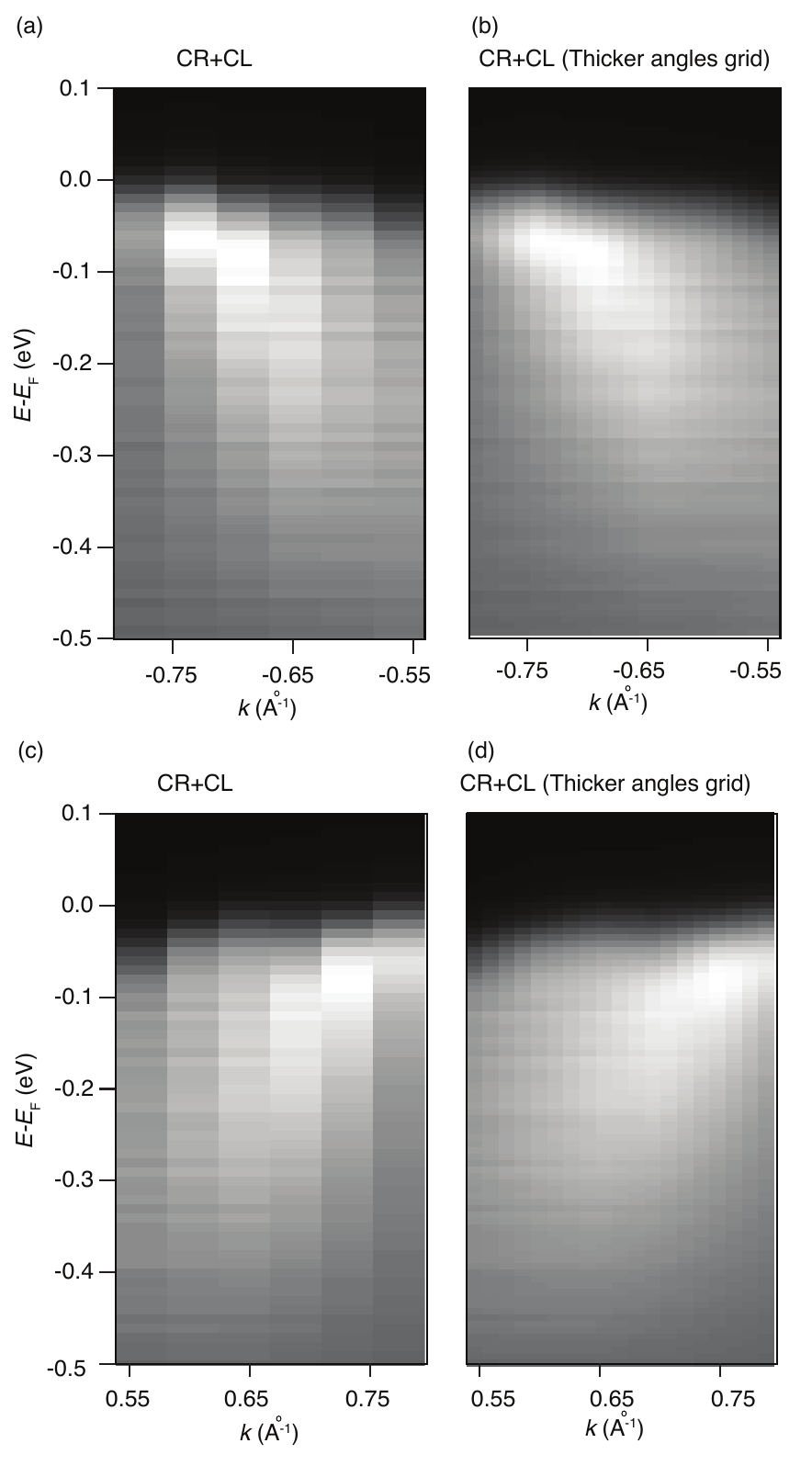}\caption{{\bf Example of some ARPES from the spin-detector} {\bf a} Energy dispersions as a function of negative $k$-values obtained by using the spin-detector only in a course alignment scan. {\bf b} Same as {\bf a} but interpolated for a thicker angular grid to see the bands better. {\bf c} Energy dispersions as a function of positive $k$-values obtained by using the spin-detector only in a course alignment scan. {\bf d} Same as {\bf c} but interpolated for a thicker angular grid to see the bands better.}\label{SpinK}
\end{figure*}

\noindent{\bf{Calibrating the V-LEED}}\\
Within the uncertainty of the instrument (1$^{\circ}$ integration region), the V-LEED has been calibrated by acquiring spin EDCs at various angles, both positive and negative values, for the sample. This is done for both spin species and with the used light polarizations. In the present case, for consistency, we did this with circularly polarized light (both left and right-handed). After, by summing up both circular polarizations and both spin species, we can reconstruct as in Fig. \ref{SpinK} the ARPES spectra. This procedure was done by using only the spin-detector to access directly the states probed and be sure that when selecting the angular values on the deflectors we probe effectively the selected state.

\noindent{\bf{Discussion about uncertainties and additional calibration}}\\
\FM{For evaluating the uncertainty, we used a controlled and known sample with no asymmetries in the dichroic signal, as in our previous work (see Ref. \cite{DiSante_2023_M}). We used a kagome lattice because at the $\Gamma$ point there is a well-defined energy gap opened due to the action of spin-orbit coupling. In addition, in this point, the bands are spin-degenerate. Plus, the system is non-magnetic. This allowed us to check the asymmetry not only in the circular dichroism signal, but also in the spin-resolved circular dichroism. We estimated the uncertainty to be approximately 10$\%$ on the residual of the dichroism. Note that this is also consistent with what is obtained by standard ARPES in our setup: at the centre of the Brillouin zone, the difference between circular right and circular left polarized spectra (each spectrum was normalized by its own maximum intensity beforehand) is indeed 10$\%$.}
\\
\begin{center}
{\bf{\Large{Supplementary Information}}}  
\end{center}

\section{Model and broken symmetry phases with chiral currents}

We consider a two-dimensional system whose electronic states at the Fermi level are described by $d$-orbitals belonging to the $t_{2g}$ manifold ($d_{xy}$,$d_{xz}$,$d_{yz}$) and can accommodate electronic phases which are marked by non-trivial loop currents.

The model Hamiltonian for the surface of the Sr$_2$RuO$_4$ is compatible with the $C_{4v}$ point group symmetry. 
For the $t_{2g}$ manifold of the $d_a$ orbitals we introduce the components of the local orbital angular momentum. 
For the spin space, hereafter, we employ the Pauli matrices $\hat{s}_{j=x,y,z}$.
Assuming that the basis of the local creation operator of electrons for the $d$-orbitals for a given spin orientation is
$\hat{d}^{\dagger}_{\bm{k}}=[c^{\dagger}_{xy,\bm{k}}, c^{\dagger}_{yz,\bm{k}},c^{\dagger}_{xz,\bm{k}}]$, we have that the components of the orbital angular momentum are given by: 
\begin{eqnarray}
\label{eq:orb}
\hat{L}_x=\begin{pmatrix} 0 & -i & 0\\
i & 0 & 0 \\
0 & 0 & 0 
\end{pmatrix}
\hat{L}_y=\begin{pmatrix} 0 & 0 & -i\\
0 & 0 & 0 \\
i & 0 & 0
\end{pmatrix}
\hat{L}_z=\begin{pmatrix} 0 & 0 & 0\\
0 & 0 & -i \\
0 & i & 0 \,.
\end{pmatrix} \nonumber
\\
\end{eqnarray}
with $\hat{L}_i$ fulfilling the usual angular momentum algebra.

Then, in the complete spin-orbital basis, given by
$\hat{C}^{\dagger}_{\bm{k}}=[c^{\dagger}_{xy,\uparrow\bm{k}}, c^{\dagger}_{yz,\uparrow\bm{k}},c^{\dagger}_{zx,\uparrow\bm{k}},c^{\dagger}_{xy,\downarrow\bm{k}}, c^{\dagger}_{yz,\downarrow\bm{k}},c^{\dagger}_{zx,\downarrow\bm{k}}]$,
the Hamiltonian can be generally expressed as 
\begin{equation}
\hat{\mathcal{H}}=\sum_{\bm{k}}\hat{C}^{\dagger}_{\bm{k}}\hat{H}(\bm{k})\hat{C}_{\bm{k}} ,
\end{equation}
with $\hat{H}({\bf k})$ given by
\begin{eqnarray}
\hat{H}&=& \sum_{\bm{k}} [(\epsilon_{xy}(k_x,k_y)-\mu)\mathbb{P}_z+ (\epsilon_{xz}(k_x,k_y)-\mu) \mathbb{P}_y + \nonumber \\&& (\epsilon_{yz}(k_x,k_y)-\mu) \mathbb{P}_x +\epsilon_{xz,yz}(k_x,k_y) (\hat{L}_x \hat{L}_y+\hat{L}_y \hat{L}_x)+ \nonumber \\ && \nonumber
\alpha_{OR} (\sin(k_x) \, \hat{L}_y -\sin(k_y) \hat{L}_x)] \hat{\sigma}_{0} + \lambda_\mathrm{SO}\hat{\bm{L}}\cdot\hat{\bm{\sigma}} \,.
\label{ham}
\end{eqnarray}

We recall that
$\mathbb{P}_a=(\hat{L}^2-2 \hat{L}_a^2)/2$ is the projector on the orbital state with a given distribution that is perpendicular to the $a$ direction (e.g. for $a=z$ the projection is on the $xy$ orbital configuration).
Here, $\mu$ is the chemical potential. The term proportional to $\epsilon_{xz,yz}(k_x,k_y)$ corresponds to the symmetry allowed hybridization processes of the $(xz,yz)$ orbitals along the $[110]$ and $[\bar{1}10]$ orientations.
The choice of the following electronic parameters reproduces with good accuracy the experimental profile of the Sr$_2$RuO$_4$ Fermi lines at the surface as measured by ARPES (as in this paper and in Ref. \onlinecite{volodya}, see e.g. Fig. 2):
\begin{eqnarray*}
\epsilon_{xy}(k_x,k_y)= && -2 t_3 [\cos(k_x)+\cos(k_y)]+\\&&-4 t_4 \cos(k_x) \cos(k_y) +\\&&- 2 t_5 [\cos(2 k_x)+ \cos(2 k_y)] \\
\epsilon_{yz}(k_x,k_y)=  && -2 t_2 \cos(k_x) -2 t_1 \cos(k_y) \\
\epsilon_{xz}(k_x,k_y)=&& -2 t_1 \cos(k_x) -2 t_2 \cos(k_y) \\
\epsilon_{xz,yz}(k_x,k_y)= && - 4 t_6 \sin(k_x) \sin(k_y) 
\end{eqnarray*}
with $t_1=0.145$, $t_2=0.016$, $t_3=0.08$, $t_4=0.039$, $t_5=0.005$, $t_6=0.001$, $\mu=0.122$, $\lambda_{\text{SO}}=0.035$, in units of eV \cite{volodya}. We assume a weak amplitude for the inversion symmetry breaking term $\alpha_{OR}\sim 0.01$ eV, since the induced splitting is below the experimental resolution. The term proportional to $\alpha_{OR}$ represents the orbital Rashba interaction \cite{Park2011,Park2012,Kim2013,Mercaldo2020} and couples the atomic angular momentum ${\bf L}$ with the crystal wave-vector ${\bf k}$ due to the breaking of inversion symmetry at the surface.

\begin{figure*}
\centering
\includegraphics[width=1.0\textwidth,angle=0,clip=true]{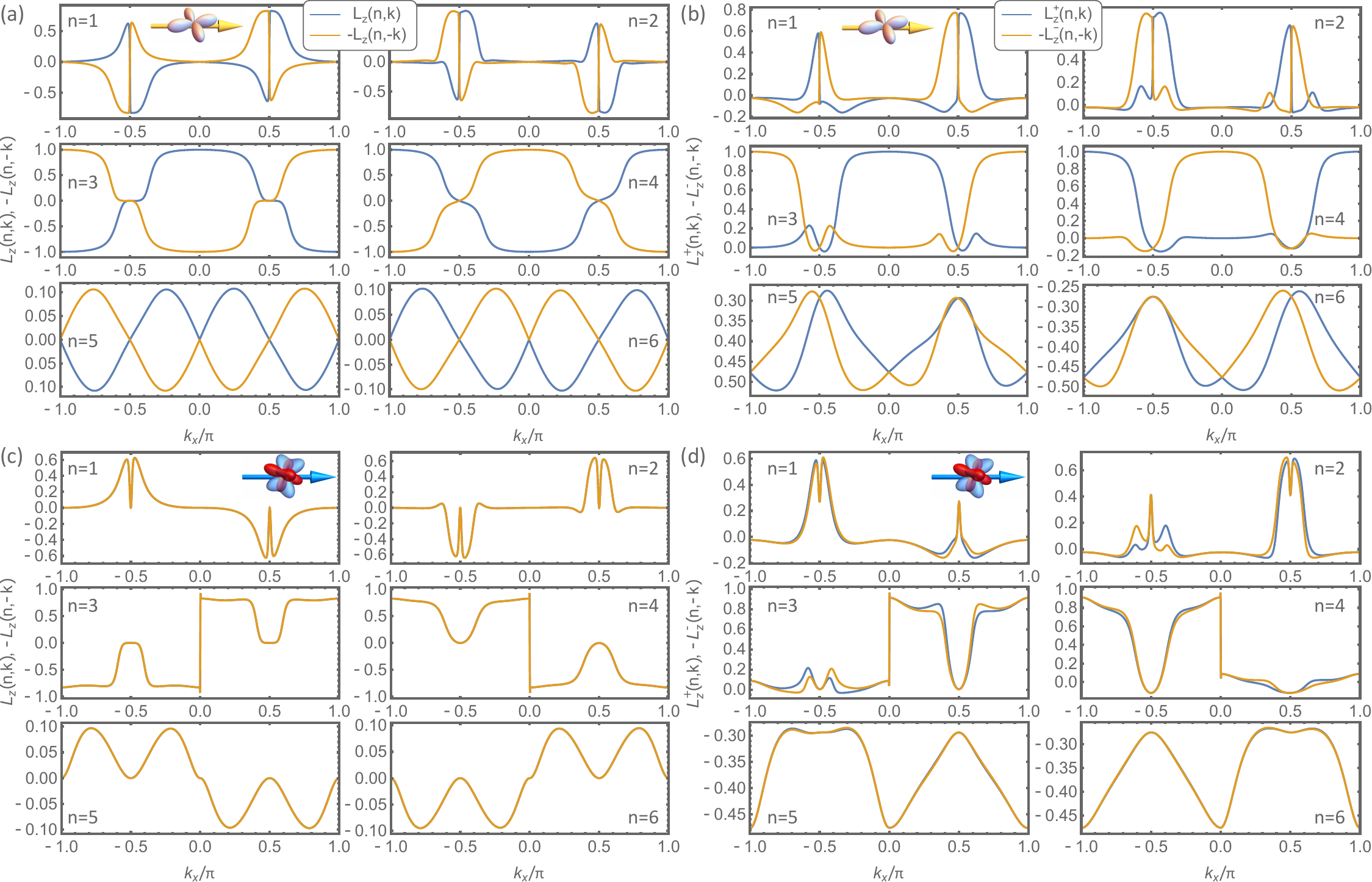}\caption{{\bf Orbital and spin-orbital textures in the presence of chiral currents.} {\bf a} Electronic phase with chiral orbital-quadrupole currents: amplitude of the orbital angular momentum ${L_z(n,k)}$ of all bands $|\psi_{n,k}\rangle$ evaluated along the ${\Gamma-X}$ direction (${L_z(n,k)}=\langle \psi_{n,k}| \hat{L}_z |\psi_{n,k}\rangle$). For clarity we plot both ${L_z(n,k)}$ and $-{L_z(n,-k)}$ for any given momentum ${k}$ to directly compare the amplitudes at opposite momenta. {\bf b} Electronic phase with chiral orbital-quadrupole currents: amplitude of the spin projected orbital angular momentum (${L^{\pm}_z(n,k)}=\langle \psi_{n,k}| (1\pm \hat{s}_z) \hat{L}_z |\psi_{n,k}\rangle$). The amplitudes of ${L_z(n,k)}$ and ${L^{\pm}_z(n,k)}$, displayed in {\bf a} and {\bf b}, do not show any symmetry and do not match at ${k}$ and $-{k}$. {\bf c} Electronic phase with chiral spin-orbital quadrupole currents with antisymmetric $L\,S$ combination:  ${L_z(n,k)}$ and ${L_z(n,-k)}$ coincide. {\bf d} Electronic phase with chiral spin-orbital quadrupole currents with antisymmetric $L\,S$ combination: spin projected orbital moment ${L^{\pm}_z(n,k)}$ at opposite momenta are unequal in amplitude.}\label{Sup4}
\end{figure*}

\MC{Then, in order to take into account the breaking of time-reversal and mirror symmetry, we consider the spin-orbital chiral current phase along the $l=[110]$ direction, as given the following term in the Hamiltonian}:
\begin{equation}
\label{eq:jso}
\hat{j}_{\text{so}}=\bm{g}_{\text{so}}\cdot(\hat{\bm{L}}\times\hat{\bm{\sigma}}) \sin k_{l} \,. 
\end{equation}
Here, we take a representative $\bm{g}_{\text{so}}$ vector with amplitude $|\bm{g}_{\text{so}}|=0.04$ eV, with generic coefficients, so that it does not point in any high-symmetry direction in the lattice. The outcome of the orbital and spin-orbital quadrupole moments within the Brillouin zone is not qualitatively affected by the choice of the coefficients.
This current term can arise by a mean-field decoupling of the Coulomb interaction for nearest-neighbor Ru centers (see Sect. IV).  
In Fig. \ref{Sup4} we report the orbital and spin-orbital textures for all the bands assuming different types of broken symmetry chiral phases marked by orbital and spin-orbital quadrupole currents. 

In order to capture the different behavior of the orbital and spin-orbital moments in the presence of chiral currents one can consider the equation of motion associated with the expectation values of the spin and orbital moments for a given eigenstate at momentum $k$.
In general, the time dependent evolution of the spin and orbital angular momentum is provided by the equation of motion $\frac{d {\bm{M}}}{d t}= \bm{\tau_M}$ with $\bm{M}=\bm{s,L}$ and ${\bm{\tau}}_M$ being the corresponding torques. For a given eigenstate $|\psi_{nk}\rangle$ of the band $n$ at momentum $k$, one can write a torque equation for the amplitudes of the spin ${\bm{s}}_{nk}=\langle \psi_{nk}| {\hat{\bm{\sigma}}}|\psi_{nk}\rangle$ and orbital ${\bm{L}_{nk}}=\langle \psi_{nk}| \hat{{\bm{L}}}|\psi_{nk}\rangle$ moments. The equation of motion reads as $\frac{d {\bm{M}}_{nk}}{d t}=\bm{\tau}^{M}_{nk}$ with $\textbf{M}=\textbf{s,L}$ and $\bm{\tau}^{M}_{nk}=i \langle \psi_{nk}| [H,{\bm{\hat{M}}}]|\psi_{nk}\rangle$. Now, since we deal with equilibrium configurations in time-independent conditions the amplitudes of the spin and orbital moments have to fulfill the relations $\tau^{M}_{nk}=0$ with $M=s,L$. These relations in turn set out the amplitude of the spin and orbital moments.

For convenience and clarity, we introduce a Cartesian reference $xyz$ and, without loss of generality, we assume that the chiral current in the broken symmetry phase flows parallel to the $x$-direction. The spin-orbital current in Eq. \eqref{eq:jso} can be then generally expressed as
\begin{eqnarray}
\hat{j}_{\text{so}}&&=[g^{x}_{so} (\hat{L}_y \hat{\sigma}_z-\hat{L}_z \hat{\sigma}_y) + g^{y}_{so} (\hat{L}_z \hat{\sigma}_x-\hat{L}_x \hat{\sigma}_z) \nonumber  \\ &&+g^{z}_{so} (\hat{L}_x \hat{\sigma}_y-\hat{L}_y \hat{\sigma}_x) ] \sin k_{x}
\end{eqnarray}
Furthermore, it is also convenient to write the Hamiltonian in the following compact form:
\begin{eqnarray}
\hat{H}=H_0+ \hat{H}_{so} + \hat{j}_{\text{so}}
\end{eqnarray}
with $H_0$ containing the kinetic and the orbital Rashba coupling terms and $\hat{H}_{so}=\lambda_{so} \hat{\bm{L}}\cdot\hat{\bm{\sigma}}$. We notice that $H_0$ preserves the vertical mirrors ($M_x$ and $M_y$) and the time-reversal symmetries.
Then, we point out that among the components of $\hat{j}_{\text{so}}$, the term related to $g^{x}_{so}$ is the source of chiral symmetry breaking, as all the mirror symmetries are broken. This is because the components of $\hat{\mathbf{L}}$ and $\hat{\bm{\sigma}}$ are pseudovectors and are perpendicular to the current flow direction. Instead, the other components, related to $g^{y}_{so}$ and $g^{z}_{so}$, preserve the $M_x$ vertical mirror symmetry. 

Let us now consider the spin and orbital amplitudes of the torque by taking for instance the corresponding orientations along the $x$-direction. 
For this configuration, the commutators of the Hamiltonian with the $x$ component of spin  and orbital angular momentum yield
\begin{eqnarray}
[\hat{H}_0,\hat{\sigma}_x] &=& 0 \\
\left[\hat{H}_{so},\hat{\sigma}_x \right] &=& i \lambda_{so} ( \hat{L}_{z} \hat{\sigma}_{y} -\hat{L}_{y} \hat{\sigma}_{z}) \equiv  i \lambda_{so}\hat{A} \\
\left[\hat{j}_{so},\hat{\sigma}_x\right] &=& i \left[ h_{x}({k_x}) \hat{B}_x +h_{y}({k_x}) \hat{B}_y +h_{z}({k_x}) \hat{B}_z ) \right] 
\end{eqnarray}
where $\hat{B}_x=\hat{L}_{y} \hat{\sigma}_{y} + \hat{L}_{z}\hat{\sigma}_{z}$, $\hat{B}_y=\hat{L}_{x} \hat{\sigma}_{y}$, $\hat{B}_z=-\hat{L}_{x} \hat{\sigma}_{z}$ and $h_i(k_x)=g_{so}^{i} \sin(k_x)$ with $i=x,y,z$. 
On the basis of these relations, the equilibrium condition for the spin torque yields:
\begin{eqnarray}
\label{eq:torqueS}
\tau^{s_x}_{nk}=\tau^{s_x,so}_{nk}+\sum_{i=x,y,z}\tau^{s_x,j^{i}_{so}}_{nk}=0 \, 
\end{eqnarray}
with 
\begin{eqnarray}
\tau^{s_x,so}_{nk}&=&-\lambda_{so}\langle \psi_{nk}| \hat{A}|\psi_{nk}\rangle \\  \tau^{s_x,j^{i}_{so}}_{nk}&=&- h_i(k_{x})\langle \psi_{nk}| \hat{B}_i|\psi_{nk}\rangle \,. 
\end{eqnarray}
We notice that in the equation for $\tau^{s_x}_{nk}$ the torque arising from the $x$ component of the spin-orbital current (i.e. $\tau^{s_x,j^{i}_{so}}_{nk}$) breaks the $M_x$ mirror symmetry. Instead, the terms related to $g_{so}^{y,z}$ are mirror-symmetric.
We also observe that in the absence of the spin-orbital current the spin torque due to the spin-orbit coupling is identically zero. This implies that the spin and orbital moments must be collinear at any $k$. On the other hand, the presence of the chiral spin-orbital currents leads to a deviation from the collinearity between $s$ and $L$.

Let us now consider the analogous equations for the orbital moment. 
For the commutators with $\hat{L}_x$ we have:
\begin{eqnarray*}
[\hat{H}_0,\hat{L}_x]&& = \\&& -i f_{y}({\bf k})\hat{L}_z + \\&& - i \sum_{a=x,y,z}(g_{ay} \{\hat{L}_a,\hat{L}_z\}-g_{zy} \{\hat{L}_a,\hat{L}_y\}\;  
\\
\left[\hat{H}_{so},\hat{L}_x \right] &=& - i \lambda_{so} \hat{A} \\
\left[\hat{j}_{so},\hat{L}_x\right] &=& i ( - h_{x}({k_x}) \hat{B}_x  + h_{y}({k_x}) \hat{C}_y +h_{z}({ k}_x) \hat{C}_z)
\end{eqnarray*}
with $\hat{C}_y=\hat{L}_{y} \hat{\sigma}_{x}$ and $\hat{C}_z=\hat{L}_{z} \hat{\sigma}_{x}$, and $[\hat{H}_0,\hat{L}_x]=i {\hat{Q}}({\bf k}) $. Here, the coefficient $f_y({\bf k})=\alpha_{OR} \sin(k_x)$ is related to the orbital Rashba coupling.
Hence, the equation for the orbital moment torque becomes:
\begin{eqnarray}
\label{eq:torqueL}
\tau^{L_x}_{nk}=\tau^{L_x,0}+\tau^{L_x,so}_{nk}+\sum_{i=x,y,z}\tau^{L_x,j^{i}_{so}}_{nk}=0 \,     
\end{eqnarray}
with 
\begin{eqnarray}
\tau^{L_x,0}_{nk}&=& - \langle \psi_{nk}| \hat{Q}({\bf k})|\psi_{nk}\rangle\\
\tau^{L_x,so}_{nk}&=& \lambda_{so}\langle \psi_{nk}| \hat{A}|\psi_{nk}\rangle \\  \tau^{L_x,j^{x}_{so}}_{nk}&=&h_x(k_{x})\langle \psi_{nk}| \hat{B}_x|\psi_{nk}\rangle \\ 
\tau^{L_x,j^{i=y,z}_{so}}_{nk}&=&-h_i(k_{x})\langle \psi_{nk}| \hat{C}_i|\psi_{nk}\rangle
\,. 
\end{eqnarray}

Now, we observe that $\tau^{s_x,so}_{nk}=-\tau^{L_x,so}_{nk}$ and $\tau^{s_x,j^{i}_{so}}_{nk}=-\tau^{L_x,j^{i}_{so}}_{nk}$. This relation is a hallmark of the phase with spin-orbital quadrupole currents. Taking into account these relations and the Eq. \eqref{eq:torqueS}  for the spin torque, one can replace $\tau^{s_x,so}_{nk}$ and $\tau^{s_x,j^{i}_{so}}_{nk}$ in the Eq. \eqref{eq:torqueL} with the combination of $\tau^{s_x,j^{a=y,z}_{so}}_{nk}$. Hence, one can deduce an equation for the orbital torque that contains only terms that are mirror symmetry compatible. This implies that the expectation value of the orbital moment has a definite parity as for mirror-symmetric systems. 

\begin{figure*}
\centering
\includegraphics[width=1.0\textwidth,angle=0,clip=true]{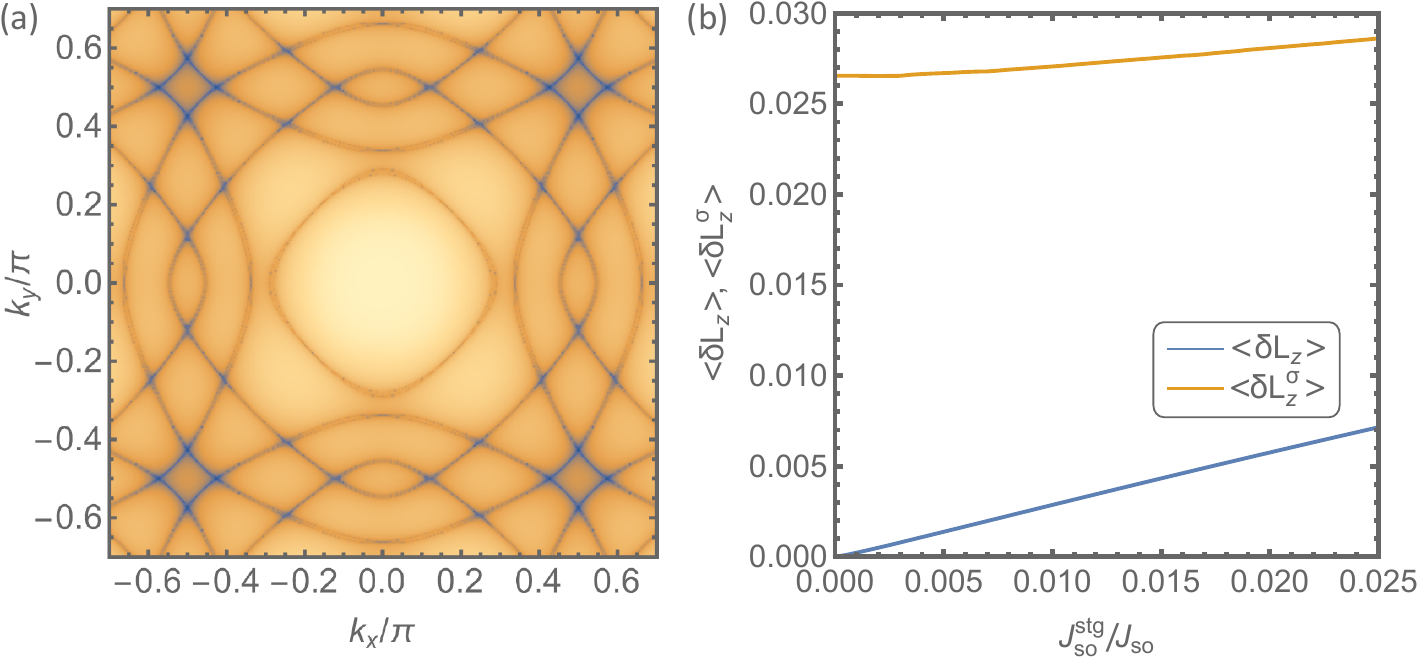}
\caption{{ {\bf Fermi surface in the presence of the staggered octahedral rotation and average asymmetry of orbital and spin-projected orbital moment.} {\bf a} Fermi surface. {\bf b} $k$-integrated and band-averaged orbital moment $\langle \delta L_z\rangle=\frac{1}{12}\sum_n\int |L_z(n,k)+L_z(n,-k)|dk$ and $\langle \delta L_z^{\sigma}\rangle=\frac{1}{12}\sum_n\int |L_z^+(n,k)+L_z^-(n,-k)|dk$ as a function of the ratio of the staggered spin-orbital quadrupole current $j^{stg}_{so}$ with respect to the homogeneous spin-orbital quadrupole current $j_{so}$ component.}}
\label{Sup6}
\end{figure*}

\begin{figure*}
\centering
\includegraphics[width=1.0\textwidth,angle=0,clip=true]{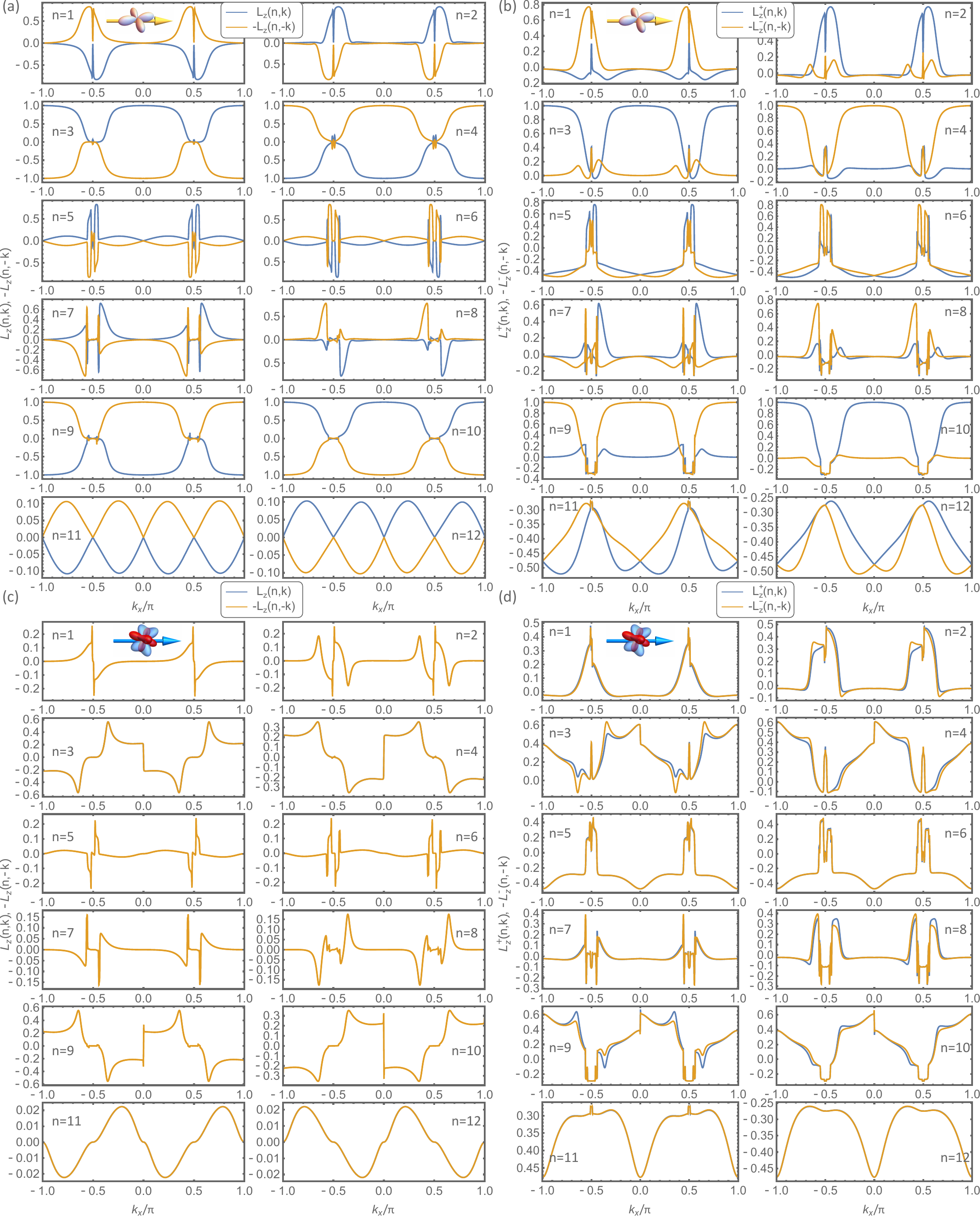}
\caption{
{\bf Orbital and spin-orbital moments in the presence of uniform chiral currents and octahedral rotation pattern.} Chiral orbital-quadrupole currents: {\bf a}   orbital ${L_z(n,k)}$ for all bands $|\psi_{n,k}\rangle$ evaluated along the $k_y=0$ direction ($\Gamma-X)$, {\bf b} spin-orbital ${L^{\pm}_z(n,k)}$. 
Chiral spin-orbital quadrupole currents: 
{\bf c}   orbital ${L_z(n,k)}$ of all bands $|\psi_{n,k}\rangle$ evaluated along the $k_y=0$ direction, {\bf d} spin-orbital ${L^{\pm}_z(n,k)}$.}\label{Sup7}
\end{figure*}

Let us now consider the case of the orbital quadrupole current:
\begin{eqnarray}
\hat{j}_{\text{o}}=&& [g^{x}_{o} (\hat{L}_y \hat{L}_z+\hat{L}_z \hat{L}_y) + g^{y}_{o} (\hat{L}_z \hat{L}_x+\nonumber \\ && \hat{L}_x \hat{L}_z)+g^{z}_{o} (\hat{L}_x \hat{L}_y+\hat{L}_y \hat{L}_x) ] \sin k_{x} \,.
\end{eqnarray}
As for the case of the spin-orbital quadrupole current, the term related to $g^{x}_{o}$ break both the horizontal ($M_z$) and vertical ($M_y$ and $M_x$) mirror symmetries, while for the other current components, we have that $g^{y}_{o}$ and $g^{z}_{o}$ preserve $M_x$. 
For this configuration one can deduce that the spin torque is not affected (i.e. $\hat{\sigma}$ commutes with $\hat{j}_o$) and that the orbital part includes all the torques due to the $g^{a=x,y,z}_{o}$ current components. Then, the resulting orbital moment has to explicitly manifest a breaking of the mirror perpendicular to the current flow as demonstrated in Fig. \ref{Sup4}. 

\subsection{Surface reconstruction}

To take into account the surface reconstruction due to the staggered octahedral rotation we double the unit cell, in order to include both sublattices of the square lattice within the unit cell. Then, the term related to the octahedral rotation can be included in the Hamiltonian as a staggered potential for the $d$-states at the Ru site of the form: 
\begin{equation}
\hat{V}_{{\rm stg}}=\eta_{\text{stg}} \hat{\tau}_z\left(\hat{L}_x\hat{L}_y+\hat{L}_y\hat{L}_x\right),
\end{equation}
where $\tau_z$ is a sublattice pseudo-spin operator given by the corresponding Pauli matrix and $\eta_{\text{stg}}$ is a characteristic energy scale of the rotation pattern. This term leads to a staggered splitting of the $(xz,yz)$ orbitals at the Ru site that arises due to the local vertical mirrors symmetry breaking resulting from the rotated octahedra. Here, for the computation of the electronic structure, we use a representative value for $\eta_{\text{stg}} = 5$ meV. Variation of the strength of $\eta_{stg}$ does not alter the qualitative outcome of the analysis. We also introduce a staggered spin-orbital current component as given by the following contribution:
\begin{equation}
\hat{j}_{\text{so}}^{\text{stg}}=
\hat{\tau}_z\left[\bm{g}_{\text{so}}^{\text{stg}}\cdot(\hat{\bm{L}}\times\hat{\bm{\sigma}}) \sin k_{l} \right]\,. 
\end{equation}
\\
Hence, regarding the chiral current broken symmetry phase, we assume an amplitude modulation of the spin-orbital current along the [110] direction so that it is given by $j^{so}_{+}=j_{so}+j_{so}^{\text{stg}}$ when connecting Ru sites in octahedra that are clockwise rotated and $j^{so}_{-}=j_{so}-j_{so}^{\text{stg}}$ for anticlockwise rotated octahedra.
\\
We have thus considered the possibility of having an amplitude modulation of the spin-orbital quadrupolar currents that follows the surface reconstruction due to the rotation of the  RuO$_6$ octahedra.  
Then, it is expected that the ground state hosts chiral spin-orbital quadrupole currents which are not spatially homogeneous, being affected by the structural change of the octahedra, that result into different current amplitudes when linking ruthenium sites with inequivalent octahedral rotation. 
We have analyzed this configuration as a function of the current amplitude unbalance associated with the two sublattices. 
For such a current pattern we find that the band resolved dichroic asymmetry in the reconstructed electronic states (Fig. \ref{Sup6} a) is not exactly zero (Fig. \ref{Sup6} b) as found for the uniform current configuration. In order to get an overall estimate of the difference between the orbital and spin-projected orbital moments as a function of the current sublattice unbalance, we have followed their integrated values (i.e. summing up all the band contributions and integrating along the $\Gamma-X$ line). The outcome is reported in Fig. \ref{Sup6} b where one can clearly see that a non-vanishing orbital moment is induced by the spatially modulated chiral currents. Nevertheless, the asymmetry for the orbital moment turns out to be significantly smaller than that related to the spin projected one. For instance, for a spatial modulated configuration of the spin-orbital current with a sublattice unbalance of around 3$\%$ we find that the integrated value of the spin-projected orbital moment is about five times larger than that of the orbital moment. 
The point here is that the spatially homogeneous chiral spin-orbital current would lead to exact zero asymmetries in the dichroic amplitude due to the balance among the torques arising from the chiral current and that one arising from the spin-orbit coupling. Now, since the spin-orbit coupling at the Ru site is homogeneous, a spatial dependent amplitude of the chiral spin-orbital current cannot be compensated and a non-vanishing amplitude of the dichroic signals can occur. Due to the surface reconstruction of the electronic states in the Sr$_2$RuO$_4$ one might argue that small deviations of the dichroic signals from zero can be consistent with the theoretical prediction when considering the orthorhombic configuration due to the staggered octahedral rotations around the $c$-axis. The main finding is that 
one can also account for an amplitude asymmetry of the dichroic signal keeping however a significantly larger spin-dichroic asymmetry as compared to the dichroic one.
\\
Finally, we have verified that for a uniform configuration of the spin-orbital quadrupole currents the asymmetry of the orbital moment at $k$ and $-k$ is identically zero irrespective of the surface reconstruction (Fig. \ref{Sup7}).

\subsection{C$_4$ rotational invariant spin-orbital chiral phases}

\begin{figure*}[!t]
\centering
\includegraphics[width=0.9\textwidth,angle=0,clip=true]{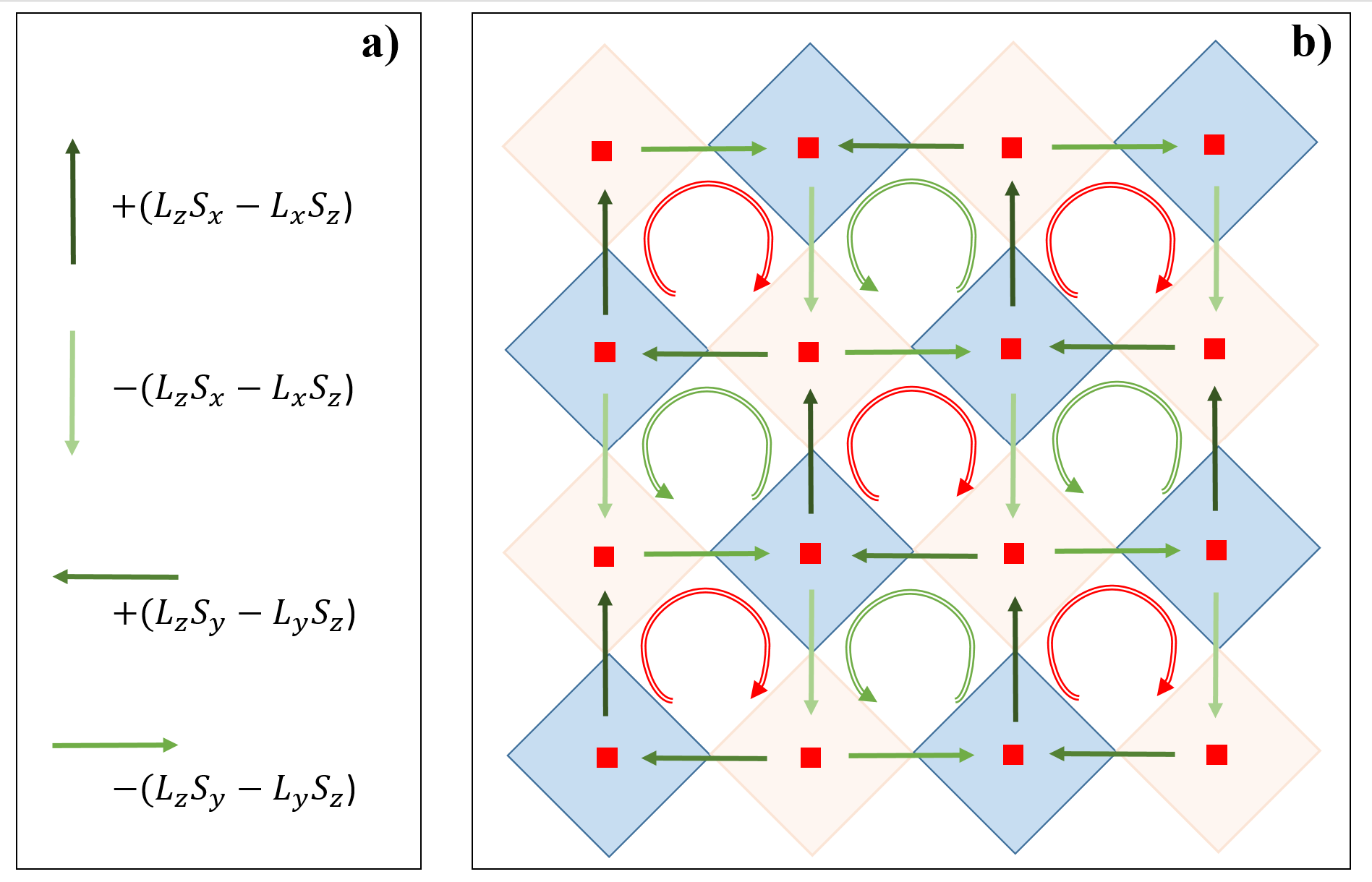}\caption{{\bf Schematic configuration of a $C_4$ rotational invariant spin-orbital chiral state.} {\bf a} The arrows indicate the current direction on a given bond associated with the corresponding spin-orbital momentum.   {\bf b} Schematic representation of a spin-orbital chiral state with $C_4$ rotation symmetry.}
\label{fig10}
\end{figure*}

We have considered states with spin-orbital chiral loop currents that preserve the C$_4$ rotational symmetry (see Fig. \ref{fig10}). These configurations lead to a symmetric dichroic amplitude and an asymmetric spin-dichroic value as related to the orbital moment ${L_z(n,k)}$ and spin projected orbital orbital moment ${L^{\pm}_z(n,k)}$ for all bands $|\psi_{n,k}\rangle$ evaluated along the $k_y=0$ direction. In particular, due to the $C_4$ symmetries the orbital moment ${L_z(n,k)}$ can be vanishing. This implies that for this type of configurations we also expect a large difference between the asymmetry in the dichroic and spin-dichroic signals.

\subsection{Canted antiferromagnetic order}

\begin{figure*}
\centering
\includegraphics[width=1.0\textwidth,angle=0,clip=true]{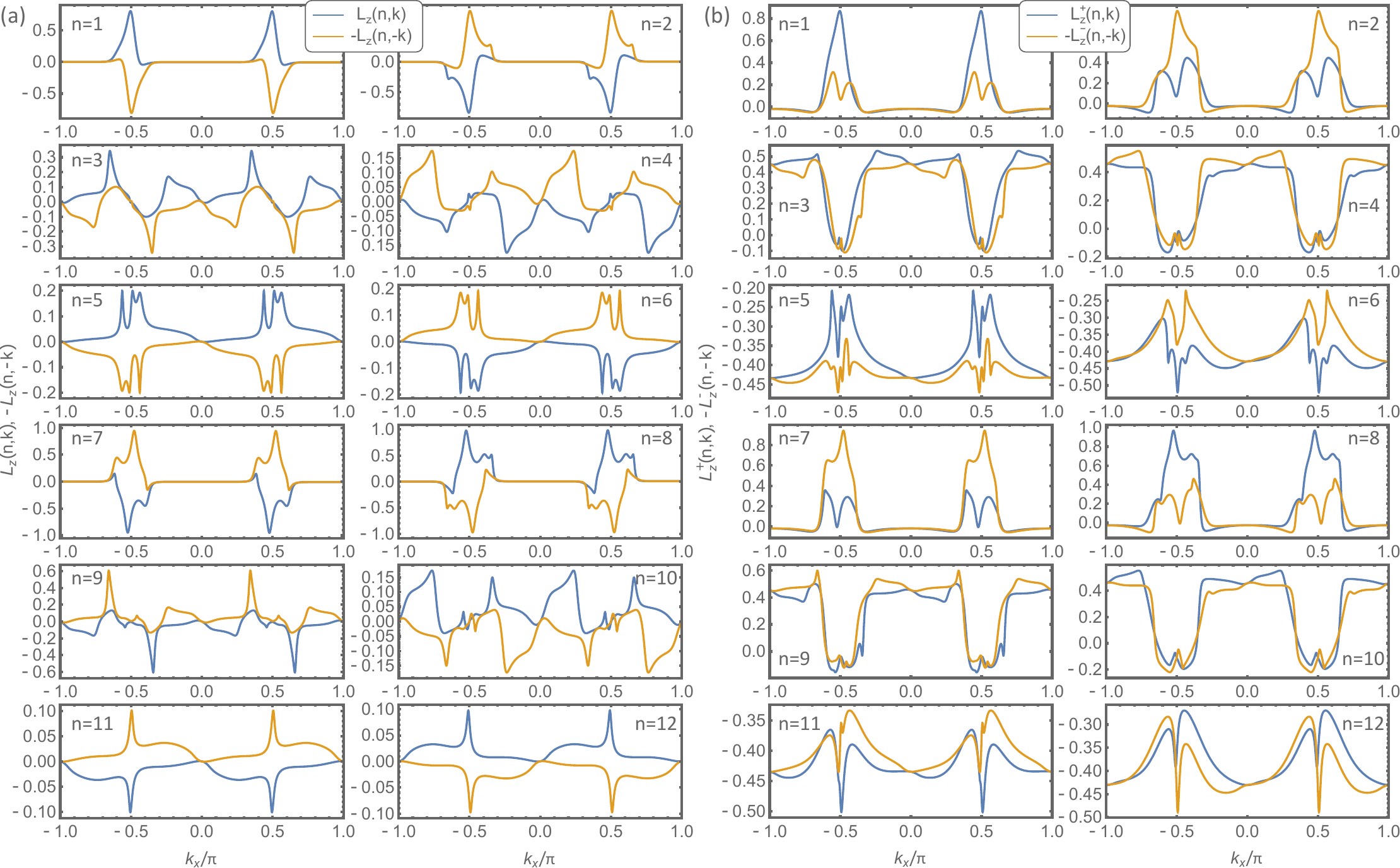}\caption{
{\bf Orbital and spin-orbital textures in the presence of canted antiferromagnetism and staggered octahedral rotation.} {\bf a}   orbital moment ${L_z(n,k)}$ and {\bf b} spin projected orbital orbital moment ${L^{\pm}_z(n,k)}$ for all bands $|\psi_{n,k}\rangle$ evaluated along the $k_y=0$ direction. 
The amplitudes of the magnetizations are: $\bm{M}_{\text{stg}}=\left(0,0,0.1\right)$ and $\bm{M}_{\text{uni}}=\left(0.02,0,0\right)$ in units of
eV$ (g\,\mu_B)^{-1}$ ($g$ is the electron $g$-factor and $\mu_B$ is the Bohr magneton.}\label{Sup8}
\end{figure*}

We consider the case of a magnetic order with a canted antiferromagnetic configuration. The broken symmetry state is introduced by an effective magnetic term in the spin channel with uniform/staggered magnetization:
\begin{equation}
H_{\text{mag}}=g \mu_B \left(\hat{\tau}_0\bm{M}_{\text{uni}}  
+ \hat{\tau}_z\bm{M}_{\text{stg}}\right)\cdot\hat{\bm{\sigma}} \,,
\end{equation}
with ${\bm{M}_{\text{uni}}}$ and ${\bm{M}_{\text{stg}}}$ being the uniform and staggered magnetization components. 
We have computed, for a representative antiferromagnetic configuration with canted moments, the orbital and spin-resolved orbital moment for all the bands crossing the Fermi level (Fig. \ref{Sup8}). As one case see from the inspection of the results in Fig. \ref{Sup8} the orbital and spin-projected orbital moments exhibit a sizable asymmetry when comparing the amplitudes at $k$ and $-k$. This is a general feature of all the magnetic phases with similar symmetry content with respect to mirror and time reversal symmetry and based on a long-range spatial order of Ru spin moments.

\section{Spin resolved circularly polarized ARPES: matrix elements and orbital angular momentum}
We report here the main steps to deduce the expressions for the dichroic and spin-dichroic ARPES transition amplitudes. To this aim, we follow the derivation reported in Ref. \cite{Park2012} and adapt it to the spin-dichroic case too.
The starting point is to consider that the circular dichroic signals probed by ARPES can be expressed through the normalized amplitude as
\begin{eqnarray}
D(k)=\sum_{\sigma}\frac{\left[I^{R}_{\sigma} - I^{L}_{\sigma}\right]}{I^{R}_{\sigma} +I^{L}_{\sigma}} \,.
\end{eqnarray}
The sum over the final state spin $\sigma$ indicates the spin-integrated nature of the detection approach. Thus, one can also introduce the spin-resolved dichroic amplitude as given by 
\begin{eqnarray}
D_s(k)=\frac{\left[I^{R}_{s} - I^{L}_{s}\right]}{I^{R}_{s} +I^{L}_{s}} \,,
\end{eqnarray}
here for convenience we assume that the spin configuration is  along the $z$-axis and are labeled by $s=+(-)$ to indicate the $\uparrow_z (\downarrow_z)$ spin states, respectively. Similar expressions can be derived for the all the other spin orientations.
\\
To calculate the amplitude of the dichroic and spin-dichroic signal one has to evaluate the transition probability $I$ for an optical excitation between the initial and final states.
The interaction with the photon is given by $H_{int}= {\bf A}\cdot {\bf p}$. The vector potentials are ${\bf A}=\frac{\epsilon_1+ i \epsilon_2}{\sqrt{2}}$ for right circularly polarized photons, and ${\bf A}^{*}=\frac{\epsilon_1- i \epsilon_2}{\sqrt{2}}$ for left circularly polarized photons. Hence, the vector ${\bf A}\times {\bf A}^{*}=-i {\bf {\hat{k}}}_{ph}$ sets out the incident photon direction.
For convenience and clarity, we report the main steps and expressions of the derivation to show the link of the ARPES intensity with the orbital and spin-projected angular momentum.
Let us consider as initial state the Bloch configuration with momentum $k$ that is given by $|\psi^{I}_{k}\rangle=\frac{1}{\sqrt{N}} \sum_{i,\alpha,\sigma} \exp[i k\cdot r_i] u_{\alpha,\sigma}(k) |i,\alpha,\sigma \rangle$, where $|u\rangle= \sum_{\alpha,\sigma} u_{\alpha,\sigma}(k) |i,\alpha,\sigma \rangle$ is the Wannier configuration centered at the site $r_i$ associated with the $t_{2g}$ orbitals labeled by $\alpha=xy,xz,yz$ and spin $\sigma$. 
A plane-wave state is assumed for the final state $|F\rangle=\int \exp[i k_F r] |r\rangle$ \cite{Damascelli2003,Moser2017} with $k_F$ being the Fermi wave vector. 
We recall that the expected value of the orbital angular momentum evaluated on the state $|\psi^{I}_{k}\rangle$ is given by 
$\langle \hat{L} \rangle=i \sum_{s} u_{s} \times u^{*}_s(k)$, with ${\bf u}_{s}=(u_{yz,s},u_{zx,s},u_{xy,s})$ and the spin projected orbital components are expressed as $\langle \hat{L} (1+s \hat{\sigma}_z) \rangle=i u_{s}(k) \times u^{*}_s(k)$, with $s=\pm$ singling out the spin up and down configuration with respect to the $z$-direction, respectively.
Following the derivation in Ref. \cite{Park2012}, one can show that the spin resolved transition probability can be generally expressed as 
\begin{eqnarray}
D_{\sigma}=\frac{({\bf A} \times {\bf A}^{*})\cdot \nabla g_{\sigma} \times \nabla g^{*}_{\sigma} }{\left[ ({\bf A}\cdot \nabla g_{\sigma}) ({\bf A}^{*}\cdot \nabla g^{*}_{\sigma})+ ({\bf A}^{*}\cdot \nabla g_{\sigma}) ({\bf A}\cdot \nabla g^{*}_{\sigma})\right]}    
\end{eqnarray}
with 
\begin{eqnarray}
g_{\sigma}(k_F)={\bf u}_{\sigma} \cdot \nabla_{k_F} f(k_F)   
\end{eqnarray}
where the gradients are respect to the Fermi wave vector $k_F$, and $f(k)$ is the Fourier transform of the part of the Wannier function that depends only on the radial distance $r-r_i$ related to the atomic center at $r_i$.
The denominator of $D$ is always positive and has a minor role in the dependence of the matrix elements from the orbital angular momentum.
The key quantity for our purposes is given by the factor $\Gamma_{s}=\nabla g_{s} \times \nabla g^{*}_{s}$. One can show \cite{Park2012} that the term $\Gamma_{s}$ can be expressed as 
\begin{eqnarray}
\Gamma_{s}= && \frac{1}{2} \varepsilon^{\alpha \beta \gamma} \left({\bf u}_{s} \times {\bf u}_{s}^{*} \right)_\alpha \nabla P_\beta f \times \nabla P_\gamma f \\ 
= && i \frac{1}{2} \varepsilon^{\alpha \beta \gamma} \langle \hat{L}_\alpha (1+s {\hat{\sigma}}_z) \rangle \nabla P_\beta f \times \nabla P_\gamma f 
\end{eqnarray}
with the vector ${\bf P}$ having the following components $(Q_{yz},Q_{zx},Q_{xy})$ with $Q_{ij}=\partial_i \partial_j$. We notice that while the structure of the orbital angular momentum is similar to that of $p$-orbitals because we are using an effective $L=1$ manifold for the $t_{2g}$ sector, the second order differential operator $Q_{ij}$ takes into account the different structure of the orbital configurations.
In a similar way, one can show that the spin integrated $\Gamma$ amplitude is given by
\begin{eqnarray}
\Gamma= i \frac{1}{2} \varepsilon^{\alpha \beta \gamma} \langle \hat{L}_\alpha \rangle \nabla P_\beta f \times \nabla P_\gamma f  \,,
\end{eqnarray}
with $\varepsilon^{\alpha \beta \gamma}$ the Levi-Civita tensor.
Then, one can see that the amplitude $\Gamma$ and thus the dichroic signal is proportional to the projected orbital angular momentum components $\langle \hat{L}_\alpha \rangle$ with respect to the incident photon direction.
Additionally, the spin-dichroic signal is proportional to the spin-projected orbital angular momentum as given $\langle \hat{L}_\alpha (1+s {\hat{\sigma}}_z) \rangle$ with $s=\pm$ for projecting spin up and down configurations, respectively.
The remaining form factors depend on the Fermi momentum and on the photon energy. Since our study aims to have a qualitative understanding of the anomalies in the transition amplitude probed by Spin CD-ARPES, their contribution does not affect the conclusions of our findings. 
\\
In the employed experimental setup the spin orientation is selected in a direction that is perpendicular to the surface (i.e. $z$) and the incident photon direction is primarily selecting the out-of-plane $z$-component of the orbital angular momentum. Hence, the dichroic and spin-dichroic signal are proportional to $\langle \hat{L}_z \rangle$ and $\langle \hat{L}_z (1\pm {\hat{\sigma}}_z) \rangle$ as considered in the manuscript.

\section{Current driven phase by Coulomb interaction}

Let us now present how the electronic current phase arises as a broken symmetry state with a nonvanishing expectation value of the current operator on the Ru-Ru bond due to the Coulomb interaction. Indeed, in order to demonstrate this point one needs to introduce the spin-orbital asymmetric  operator  
\begin{eqnarray}
    \phi^{\alpha \beta}_{\sigma,\sigma'}(l,m)=i \left( c^{\dagger}_{\alpha,\sigma}(l) c_{\beta,\sigma'}(m) -c^{\dagger}_{\beta,\sigma'}(m) c_{\alpha,\sigma}(l) \right)
\end{eqnarray}
for the $l-m$ bond between two Ru atoms with position identified by the coordinates $R_l$ and $R_m$. Here, $c_{\alpha,\sigma}(l) (c^{\dagger}_{\alpha,\sigma}(l))$ are the annihilation (creation) operators associated with an electronic state with $\alpha$ orbital and spin $\sigma$ at the atomic site $R_l$. Then, the spin and orbital dependent terms that build up the density-density inter-site Coulomb interaction $U_{lm}$ for a generic $l-m$ bond can be written in the following form
\begin{eqnarray}
U_{lm} n_{\alpha,\sigma}(l) n_{\beta,\sigma'}(m)=&&-\frac{1}{2} U_{lm} (\phi^{\alpha \beta}_{\sigma,\sigma'}(l,m))^{\dagger}  \phi^{\alpha \beta}_{\sigma,\sigma'}(l,m) \nonumber \\ && + \frac{1}{2} U_{lm} \left( n_{\alpha,\sigma}(l) + n_{\beta,\sigma'}(m) \right)    
\end{eqnarray}
where the orbital and spin resolved density operators, $n_{\alpha,\sigma}(l)$, are defined as $n_{\alpha,\sigma}(l)=c^{\dagger}_{\alpha,\sigma}(l) c_{\alpha,\sigma}(l)$. Hence, by decoupling the quartic term, one can introduce an order parameter associated with the expectation value of $\phi^{\alpha \beta}_{\sigma,\sigma'}(l,m)$ and express the interaction as 
\begin{eqnarray}
U_{lm} n_{\alpha,\sigma}(l) n_{\beta,\sigma'}(m) \sim && -\frac{1}{2} [ \langle \phi^{\alpha \beta}_{\sigma,\sigma'}(l,m)\rangle (\phi^{\alpha \beta}_{\sigma,\sigma'}(l,m))^{\dagger} + h.c. \nonumber \\ && - |\langle \phi^{\alpha \beta}_{\sigma,\sigma'}(l,m)\rangle |^2 ]
\end{eqnarray}
where the average value indicates the summation over all the electronic states weighted by the Fermi distribution function.
Taking into account the spin-orbital order parameters on the $l-m$ bond, by suitable superposition of the $\phi$ operators one can construct a bond current order parameter that is given by the expectation value of the following orbital and spin-orbital quadrupole current operators:
\begin{eqnarray}
J^{o}_{i,j}(l,m)=i \left( \vec{c}^{\,\,\dagger}(l) \hat{L}_{i} \hat{L}_{j} \vec{c}\,(m) -h.c. \right) \nonumber \\
J^{so}_{i,j}(l,m)=i \left( \vec{c}^{\,\,\dagger}(l) \hat{L}_{i} \hat{S}_{j} \vec{c}\,(m) -h.c. \right) 
\end{eqnarray}
with $\vec{c}^{\,\,\dagger}(l)=\left[c^{\dagger}_{xy,\uparrow}(l),c^{\dagger}_{yz,\uparrow}(l),c^{\dagger}_{zx,\uparrow}(l),c^{\dagger}_{xy,\downarrow}(l),c^{\dagger}_{yz,\downarrow}(l),c^{\dagger}_{zx,\downarrow}(l) \right]$ and $i,j\in\{x,y,z\}$.


%


\noindent{\bf{\large Acknowledgements}}\\
F.M. greatly acknowledges the SoE action of PNRR, number SOE\_0000068. M.C., R.F., M.T.M., and A.V. acknowledge support from the EU’s Horizon 2020 research and innovation program under Grant Agreement No. 964398 (SUPERGATE). M.C. and G.P. acknowledge financial support from PNRR MUR project PE0000023-NQSTI.  M.C., R.F., A.G., and A.V. acknowledge partial support by the Italian Ministry of Foreign Affairs and International Cooperation, grant number KR23GR06. W.B. acknowledges support by Narodowe Centrum Nauki (NCN, National Science Centre, Poland) Project No. 2019/34/E/ST3/00404 and partial support by the Foundation for Polish Science through the IRA Programme co-financed by EU within SG OP. This work was performed in the framework of the Nanoscience Foundry and Fine Analysis (NFFA-MUR Italy) facility. J.A.M acknowledges financial support from DanScatt (7129-00011B).

\end{document}